\RequirePackage{fix-cm}
\documentclass{article}

\usepackage{arxiv}
\usepackage[tight,footnotesize]{subfigure}
\usepackage{graphicx}
\usepackage{multicol}
\usepackage{amssymb,amsmath}
\usepackage[]{algorithm2e}
\usepackage{color}
\usepackage{soul}
\usepackage[normalem]{ulem}
\graphicspath{{figures/}}
\usepackage{pgffor}
\usepackage{hyperref}    

\begin{document}

\title{User Attention and Behaviour in Virtual Reality Art Encounter}


\author{ \href{https://orcid.org/0000-0003-1931-7959}{\includegraphics[scale=0.06]{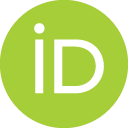}\hspace{1mm}Mu~Mu}\\
	University of Northampton\\
	Northampton, United Kingdom \\
	\texttt{mumu@ieee.org} \\
	\And
	Murtada~Dohan\\
	University of Northampton\\
	Northampton, United Kingdom \\
	\texttt{murtada.dohan@northampton.ac.uk} \\
	\And
	Alison~Goodyear \\
	Alison Goodyear Artist\\
	Bedford, United Kingdom \\
	\texttt{alisongoodyear@me.com} \\
	\And
	Gary~Hill \\
	University of Northampton\\
	Northampton, United Kingdom \\
	\texttt{gary.hill@northampton.ac.uk} \\
	\And
	Cleyon~Johns \\
	University of Northampton\\
	Northampton, United Kingdom \\
	\texttt{cleyon.johns@northampton.ac.uk} \\
	\And
	Andreas~Mauthe \\
	Universitat Koblenz-Landau, \\
	Koblenz, Germany\\
	\texttt{mauthe@uni-koblenz.de} \\
	}
	

\maketitle

\begin{abstract}
With the proliferation of consumer virtual reality (VR) headsets and creative tools, content creators have started to experiment with new forms of interactive audience experience using immersive media. Understanding user attention and behaviours in virtual environment can greatly inform creative processes in VR. We developed an abstract VR painting and an experimentation system to study audience encounters through eye gaze and movement tracking. The data from a user experiment with 35 participants reveal a range of user activity patterns in art exploration. Deep learning models are used to study the connections between behavioural data and audience background. New integrated methods to visualise user attention as part of the artwork are also developed as a feedback loop to the content creator.


\keywords{virtual reality \and VR abstract painting \and user attention \and eye-tracking \and machine learning}
\end{abstract}

\section{Introduction}
\label{sec:intro}



With the proliferation of consumer virtual reality (VR) headsets and readily available creative tools, it is evident that more content creators and artists are experimenting with alternate interactive audience experiences using immersive media \cite{goodyear2019abstract,10.1145/3240508.3243718}. Creating content using new media brings new opportunities, which historically, is an approach long-used in the field of fine art. With VR, visual art artists can enjoy ``unrestricted" space for creative work while exploiting features that are not possible in the physical world \cite{grau2003virtual}. However, employing new media and alternate reality in fine art requires new understanding of how audiences interact with and experience new forms of content to inform a creative process. For the visual arts, this means establishing a better understanding of how viewers approach an artwork and navigate through a virtual environment. The research on user attention in VR encounter can lead to new developments within art practice where user activity data and computer programming can serve as a dynamic tool for innovation \cite{10.1145/3317697.3325118}. 

In order to further explore the topic of content creation in alternate reality, a group of computer science researchers, 3D games designers, and a fine art artist collaborated on a VR art project. Goodyear is experienced in using Google Tilt Brush to create experimental abstract VR paintings. An experimentation system is developed using Unity games engine in order to install any artwork and enable user navigation. With a combination of hardware and software functions, the system is equipped with multi-modal eye gaze and body movement tracking capabilities to capture user interactions with the artwork. A user experiment was conducted, where 35 participants explored an abstract VR painting according to their own preference. The participants' eye-gaze, body movements, and voice comments were recorded while they freely moved within a 3 by 4 metre physical space. 
The majority of participants approached the virtual works with confidence, to become immersed within the artwork. Many participants claimed that the new form of media encouraged an interest in not only these abstract VR paintings, but also abstract art in general.
The quantitative experimental data showed distinctive activity patterns which reflect different preferences in moving locations, body positions, and viewing angles.  
In order to model and predict user preferences, deep-learning techniques were employed to study the connections between users behavioural data and the audience background/previous experience. To enhance the feedback loop from user experiments to content creators, new ways to visualise user attention in virtual environments were also piloted. The results can be particularly useful for artists who are looking to adapt their art practices based on audience experiences and for the creative industry to develop new applications. It is felt that a tracking and data analysis engine, paired with games development engines have the possibility to become a new tool to augment art creation process, echoing how AI assists engineering and computer science designs.

Main contributions of this paper include:
\begin{itemize}
\item A system that supports systematic studies of interactive VR artwork.
\item An experiment with 35 participants to study of how people interact with an abstract VR painting using eye-gaze and body movement tracking.
\item New integrated methods for visualising user attention in 3D virtual environments for VR content creators to evaluate how the content has been perceived.
\item The use of deep learning models to analyse multi-dimensional user behavioural data to support future development of personalised encounters.
\end{itemize}

The remainder of this paper is organised as follows. Section \ref{sec:background} discusses the background and related work in VR art and behavioural tracking in VR. Section \ref{sec:experiment} introduces the authors' VR artwork, experimentation system, and the user experiment. Data analysis and modelling are discussed in Section \ref{sec:data} and Section \ref{sec:modelling}. Section \ref{sec:visualisation} explores new methods for visualising user attention, whilst Section \ref{sec:discussions} includes discussions. Section \ref{sec:conclusions} concludes the paper.

\section{Related Work}
\label{sec:background}
\subsection{Virtual Reality Art}

There is an increasing adoption of alternate reality platforms by content creators and visual artists worldwide. Blortasia is an abstract art world in the sky where viewers fly freely through a surreal maze of evolving sculptures \cite{mack2017blortasia}. 
Authors believe the exploration through art and nature reduces stress, anxiety and inflammation, and has positive effects on attitude, behaviour, and well-being. Hayes, et al. created a virtual replication of an actual art museum with features such as gaze-based main menu interaction, hotspot interaction, and zooming/movement in a 360 degree space. Authors suggested that allowing viewers to look around as they please and focus their attention on the interaction happening between the artwork and the room is something that can’t be replicated easily \cite{10.1145/3281505.3281620}. In \cite{8611753}, Battisti, et al. presented a framework for a virtual museum based on the use of HTC VIVE. The system allows moving in the virtual space via controllers as well as walking. A subjective experiment showed that VR, when used in a cultural heritage scenario, requires that the system should be designed and implemented by relying on multi-disciplinary competences such as arts and computer science.

Zhou conducted a systematic, theoretical approach to study the issues involved in VR Art Museum and inform art museum professionals in their decision making process \cite{10.1145/3366344.3366441}. The discussions revolved around accessibility, enrichment, interaction, presence, and content. Parker et al, reported on a project designed to address some concerns in adopting VR at Anise Gallery in London with focused on multi-sensory. The research explored how the inclusion of VR might alter the practice of people watching and whether the incorporation of VR might produce a qualitatively different experience of the art museum as a shared social space \cite{Parker2020}. In \cite{raz2019virtual} Raz observed how VR is increasingly adopted by diverse artists and attains growing recognition at film festivals and argues that VR is endowed with immersive affordance, which qualitatively differ from those of any other art media. 

In February 2020 we hosted \emph{Paint Park}, an experimental installation of immersive virtual and physical paintings based at MK Gallery Project Space in the UK and created by artist Goodyear \cite{paintpark}. The installation explored the possibilities of ``phygital" painting, where digital and physical painting intersect. The works used physical materials combined with new technologies in an attempt to test what it is we understand as painting today to consider what could be best described as painting as place. 

The recent COVID-19 outbreak has greatly impacted the art and cultural sector in many parts of the world. Many galleries and museums are closed indefinitely while exhibitions and auctions are postponed or cancelled. This triggered a new wave to explore alternative digital spaces with online and VR exhibitions \cite{theguardian2020}. Now physical spaces are no longer the priority, the cultural sector is rushing to adapt events, exhibitions and experiences for an entirely digital-first audience. In America, Art Institute of Chicago and Smithsonian are among institutions that have embraced VR and taken on new significance as the lock-down deepens \cite{grant2020}.

\subsection{Eye Gaze and Behaviour Tracking}

One of the first use cases for eye gaze tracking in VR was Dynamic Foveated Rendering (DFR), which allowed VR applications to prioritise resource usage on the foveal region where a user is looking \cite{tobii}. This helped reducing the volume of heavy rendering for complex scenes \cite{10.1145/2980179.2980246}, enhancing the image quality or improving the frame rate \cite{tobii}. Chen et al. used infrared sensors and cameras to detect and robustly reconstruct 3D facial expressions and eye gaze in real time to enhance bidirectional immersive VR communication. The research was based on the argument that users should be able to interact with the virtual world using facial expressions and eye gaze, in addition to traditional means of interaction \cite{8446494}.

In \cite{10.1145/3225153.3225157} Pfeil, et al. studied humans eye-head coordination in VR compared to physical reality. The research showed that users statistically move their heads more often when viewing stimuli in VR. 
Nonverbal cues, especially eye gaze, plays an important role in our daily communication as an indicator of interest and as a method to convey information to another party \cite{10.1145/3281505.3281587}. Kevin et al. presented a simulation of human eye gaze in VR to improve immersion of interaction between user and virtual agent. A gaze aware virtual agent was capable of reacting towards the player's gaze to simulate real human-to-human communication in VR environment \cite{10.1145/3281505.3281587}. 
Marwecki, et al. presented Mise-Unseen \cite{10.1145/3332165.3347919}, a software system that applied scene changes covertly inside the user's field of view. Gaze tracking has also been used to create models of user attention, intention, and spatial memory to determine the best strategies for making changes in scenes in VR applications \cite{10.1145/3332165.3347919}.

Embedded eye trackers were used to enable better teacher-guided VR applications since eye tracking could provide insights into student’s activities and behaviour patterns \cite{10.1145/3357251.3358752}. The work presented several techniques to visualise eye-gaze data of the students to help a teacher gauge student attention level. Pfeuffer et al. investigated body motion as behavioural biometrics for virtual reality to identify a user in the context of authentication or to adapt the VR environment to users' preferences. Authors carried out a user study where participants perform controlled VR tasks including pointing, grabbing, walking, and typing while the system monitoring their head, hand, and eye motion data. Classification methods were used to associate behaviour data to users \cite{10.1145/3290605.3300340}. Furthermore, avatars are commonly used to represent attendees in social VR applications. Body tracking has been used to animate the motions of the avatar based on the body movements of human controllers \cite{caserman2019real}. In \cite{lopez2019investigating} full visuomotor synchrony is achieved using wearable trackers to study implicit gender bias and embodiment in VR. 

\section{VR Art Experimentation System and User Experiment}
\label{sec:experiment}

The main purpose of the research described in this paper is: 1) to study how audiences interact with abstract VR paintings. This includes how they change their location by walking to explore different parts of the artwork or to change their viewing perspective, how they split their attention between thousands of brushstrokes, whether they walk into the artwork to have a more immersive experience, whether they reach out in an attempt to touch, and how they verbalise to their experience. 2) to visualise user interactions in VR as a means to provide user experience feedback to VR visual artists and HCI designers. 3) to model eye-gaze and movement data to enable a better understanding of human attention and personal experiences in VR.

\begin{figure}[!htbp]
\centering
  \includegraphics[width=0.95\textwidth]{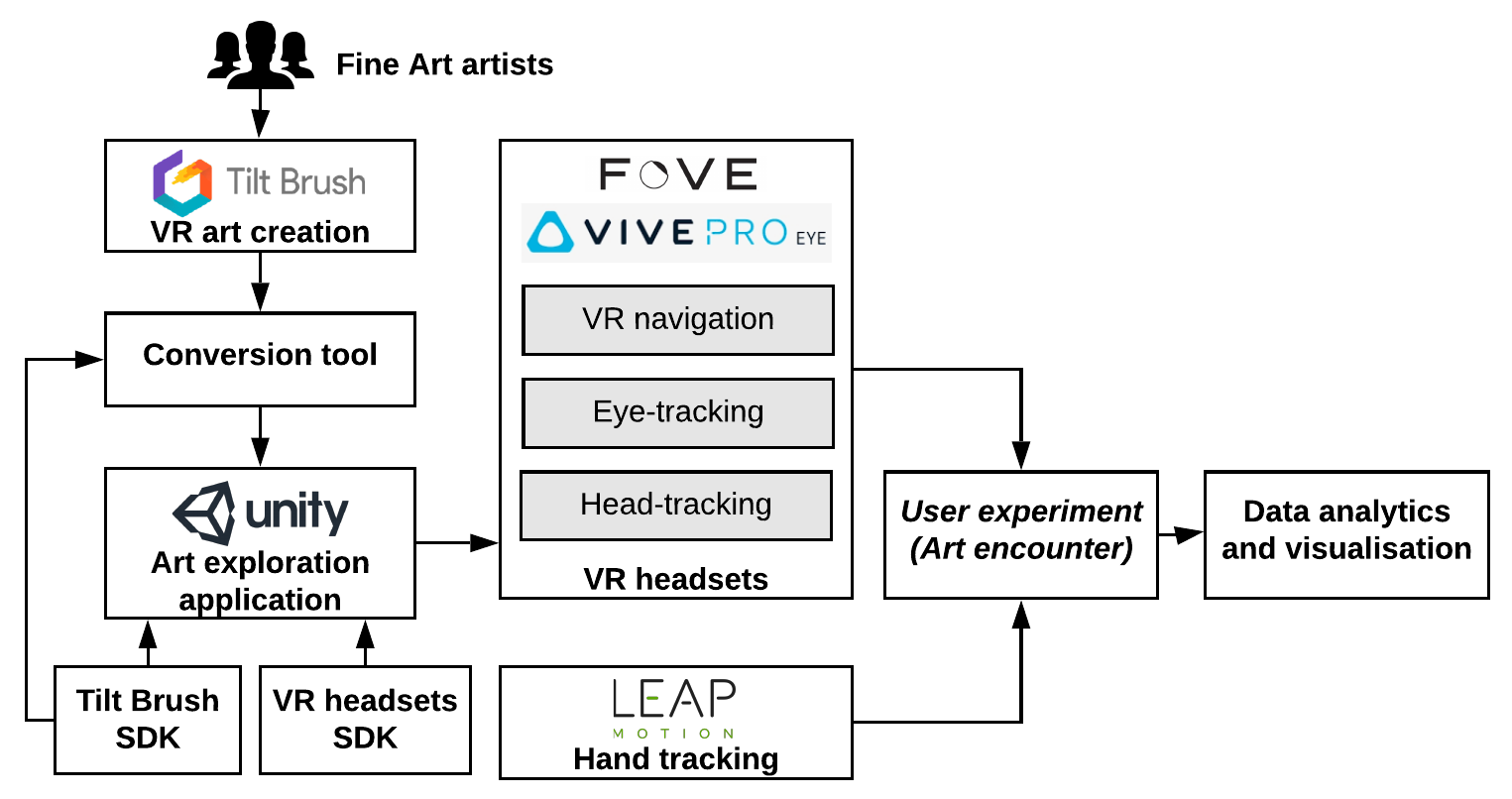}
\caption{Experimentation system diagram}
\label{fig:system}      
\end{figure}

Figure \ref{fig:system} depicts the system tailored to support the research. 
The artwork is created in Google Tilt Brush and ported to Unity, a games engine, using an in-house conversion tool. Then an art exploration application was created in the format of a VR game allowing user navigation and activity tracking. VIVE Pro Eye \cite{viveproeye} was chosen as the reference VR headset for the user experiment while the system also supports FOVE 0 \cite{fove0}. VIVE Pro Eye has built-in eye-tracking capabilities (Tobi-based) and head position tracking using base stations. An additional real-time application was developed to extract, filter and store eye gaze and behavioural data in a database. Hand tracking was also incorporated by attaching a Leap Motion onto the VR headset though the analysis of hand movement data is outside the scope of this paper. 

\subsection{Abstract VR painting}

\begin{figure}[!htbp]
\centering
    \subfigure[View 1]  {   \label{fig:paint1} \includegraphics[width=0.7\columnwidth]{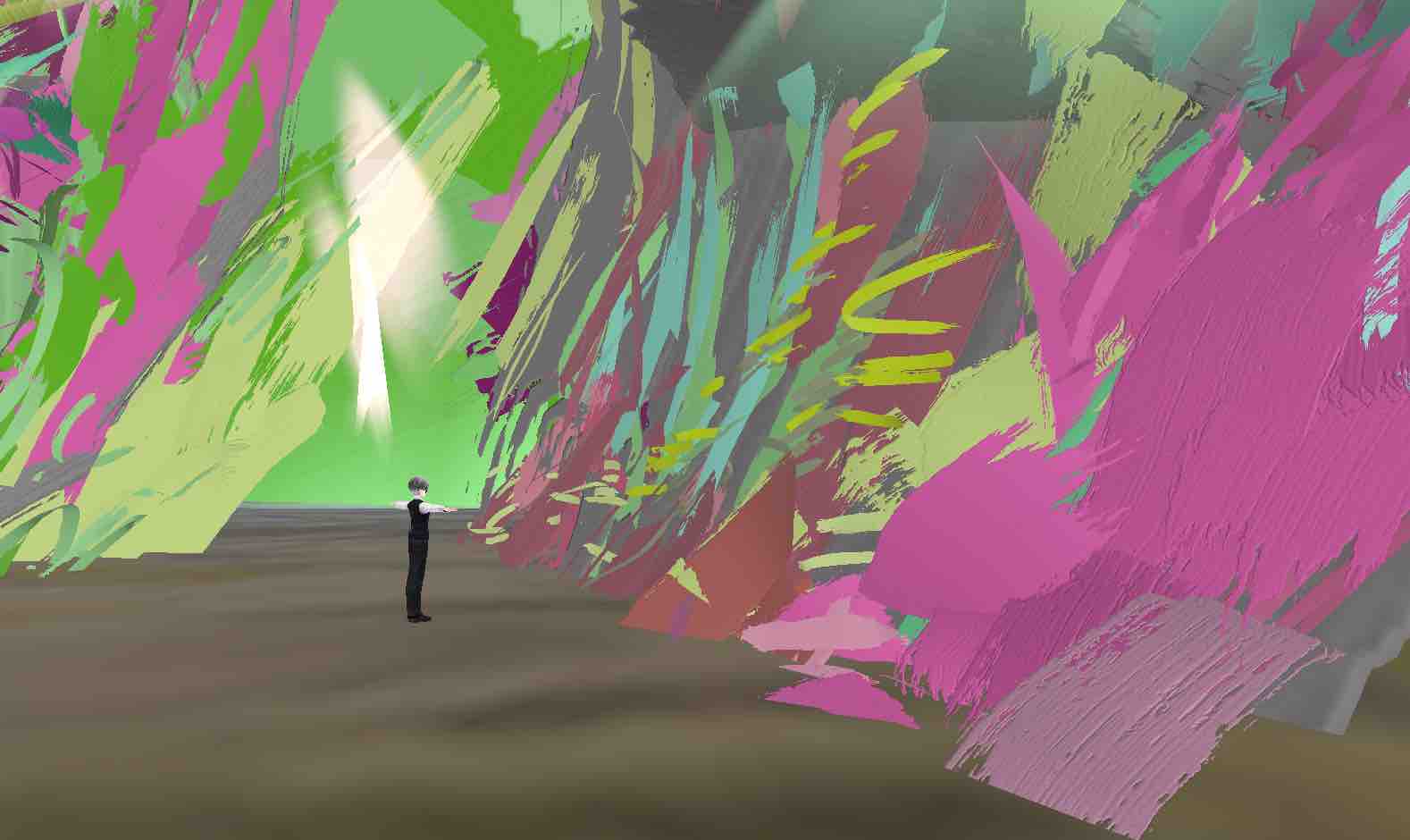} }
    \subfigure[View 2]  {   \label{fig:paint2} \includegraphics[width=0.7\columnwidth]{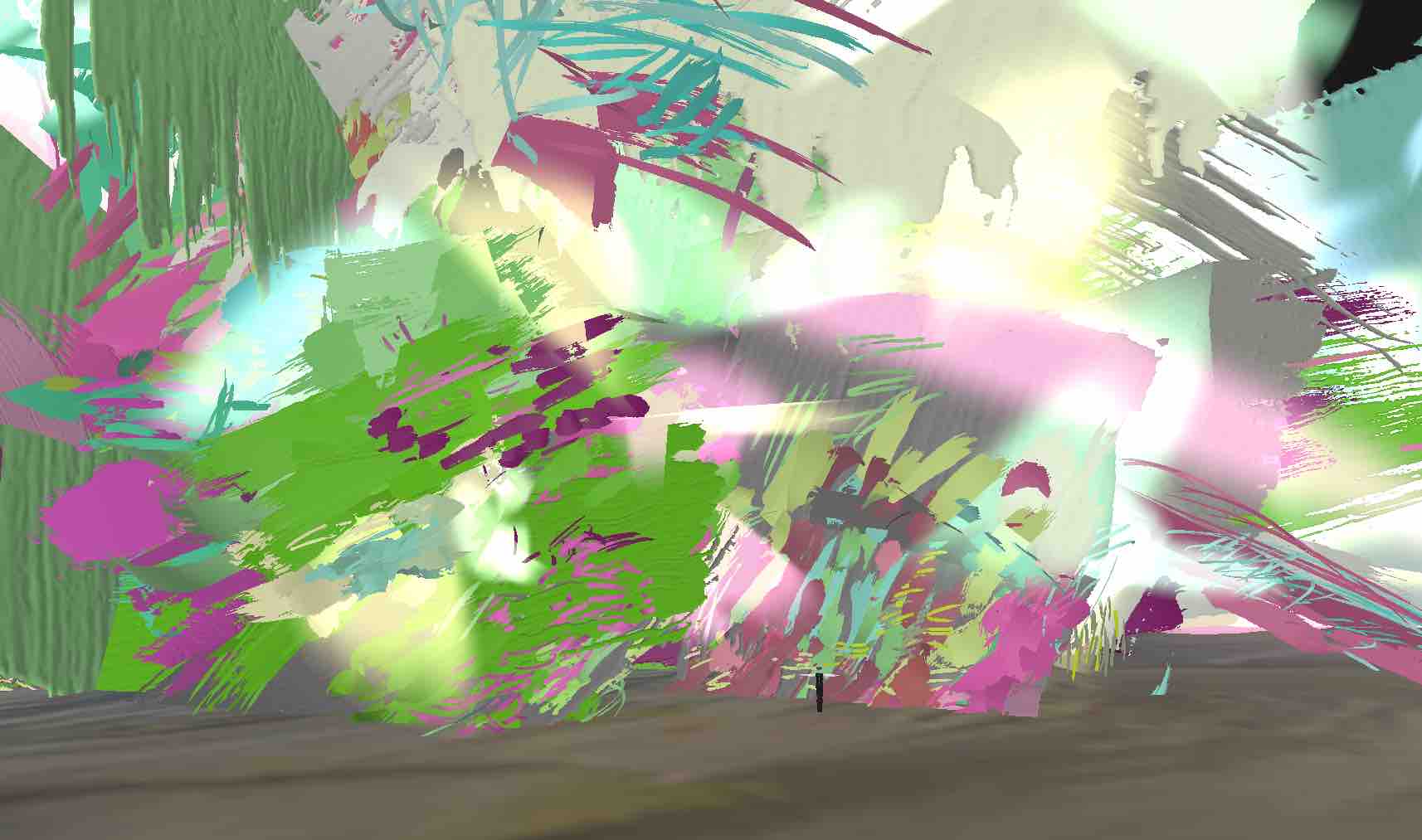} }
    \caption{Abstract VR painting (with an added human-size virtual character to indicate scale)}    \label{fig:paint}
\end{figure}

The abstract VR painting ``Caverna Coelo" (Figure \ref{fig:paint}) used for this research was created by author and artist Goodyear who was developing skills in the new medium as an abstract VR painter. 
This work draws from Goodyears more-established physical painting process, which is principally a reflection on the painting process i.e. painting about the process of painting (that is the making and the thinking behind it) \cite{alig}. To achieve this Goodyear used documentation of the painting process within the painting and in most cases this involved using their paper-based paint palettes. This process was continued in Google Tilt Brush, where an image was imported into the paint palette followed by extrapolation from it. As seen in this work, Goodear experimented with using physical palette imagery as a type of floor for the artworks and also as a backdrop. These works are described as ``phygital", where the physical and digital were combined. This term, around since the turn of the century, was originally found in marketing to describe how customers in banking would interact with online platforms. Now it has been appropriated into many different fields, including the art world.

``Caverna Coelo" uses both physical and digital painting elements and contains areas of concentrated painting marks and hidden pockets of almost isolated space, which Goodyear describes as painting caves. When navigating deep into the work through the layers within the work, these spaces are revealed, often unexpectedly, to the viewer. This approach, as the following sections will go on to discuss, shows promising signs of encouraging a virtual exploration.

\subsection{Exploration of an artwork using a games engine based application}

The artwork was ported into the Unity games engine to enable customised user navigation and behavioural tracking. While Tilt Brush allows VR creations to be exported directly in various formats such as FBX, all brushstrokes of the same type such as \emph{``Wet Paint''} are grouped as one mesh object (even when the brushes are not connected) to make the handling of artwork more efficient. This means the system would only be able to tell whether a certain type of brushstrokes was looked at. As the authors were interested in user attention at individual brushstroke level, a conversion tool has been developed based on Tilt Brush Unity Toolkit \cite{tiltbrushtoolkit} to retain brushstrokes as independent mesh objects using a unique identifier with the naming scheme: \emph{brushtype\_startingcolour\_artworkref\_seq}. For instance, an oil paint brushstroke from the VR painting \emph{peacock} with a starting colour 4278853398 (RGBA) and a unique ID of 5251 would be identified as \textit{OilPaint\_4278853398\_peacock\_5251}. The conversion tool can export the artwork as a single FBX file with brushstrokes separated internally or as separate FBX files while each holds one brushstroke. The tool also outputs the metadata of all brushstrokes as a JSON file for data analysis purposes.

The artwork was then correctly scaled in Unity by the artist and the elevation of the camera was set to mimic the artist's view when the painting was created. Various environmental configurations such as lighting were made to match how the painting was felt in Tilt Brush. Due to the scale and complexity of the artwork, occlusion culling was use to maximise the scene rendering performance and maintain a high frame-rate for the improved viewing experience. For this experiment, participants were able to explore the artwork by walking freely within a set physical area. Therefore the navigation was enabled by mapping any physical movements (walking, crouching, leaning, head turning, etc.) to camera positions and movements in Unity like a first-person computer game. The settings also allowed participants to walk through brushstrokes. 

\subsection{Eye Gaze and Behaviour Tacking}

The VIVE Pro Eye is an advanced VR headset equipped with an embedded eye tracking sub-system. It is built to meet the requirements of the most perceptive commercial clients and academic researchers since it provides research-grade data. 
VIVE Pro Eye comes with a compatible SDK that supports various games engines as well as data stream for research purposes. 

Leap motion was added to the headset in order to extend the tracking features. The SDKs of both devices work synchronously within our Unity viewing application. The idea behind this was to introduce a unique tracker approach that synchronises eye, head, body, and hand movements at the same time. Leap motion was used to provide data of hand movements and to simulate these movements in the game thus increasing content reliability from a player's perspective. 
The system also tracked eye blinking events to validate or bridge eye gaze activities. This was particularly important to identify long gaze events, which can be wrongly segmented due to eye blinking. 


VIVE Pro Eye SDK provides two types of eye tracking data, local and world space. Local space data provides eye orientation relative to the headset and is not associated with head orientation. World space data was used to  take head orientation into consideration when computing eye orientation. World space data provided the raw data (eye\_x, eye\_y, eye\_z) as a combination of both left and right eye coordinates on a unit sphere. Any tracking data marked as ``invalid" by the SDK is discarded.

The data captured via the VR headset included both head position and eye direction, as it provided coordinates of both head position (head\_x, head\_y, head\_z) and eye origin (eye\_x, eye\_y, eye\_z). Head coordinates are correlated to real world measurements. Meaning that, head\_x, head\_z coordinates refer to width and depth of the inside walk zone whereas, head\_y refers to the height of the head. While, eye origin refers to a point in the world space where the eye was looking at and its coordinates help to generate the direction vector in the game that indicated the final position of what users have seen. In addition, Unity provided the player position along with head rotation that represented the direction of the head in the field using four dimensions (rotation\_x, rotation\_y, rotation\_z, rotation\_w) named quaternion. Player position (player\_x, player\_y, player\_z) represented the location of the participants within the virtual environment. Finally, a timestamp was generated for each data frame to enable sequencing and validation of the raw data.


\subsection{User Experiment}

\begin{figure}[!htbp]
\centering
  \includegraphics[width=0.8\textwidth]{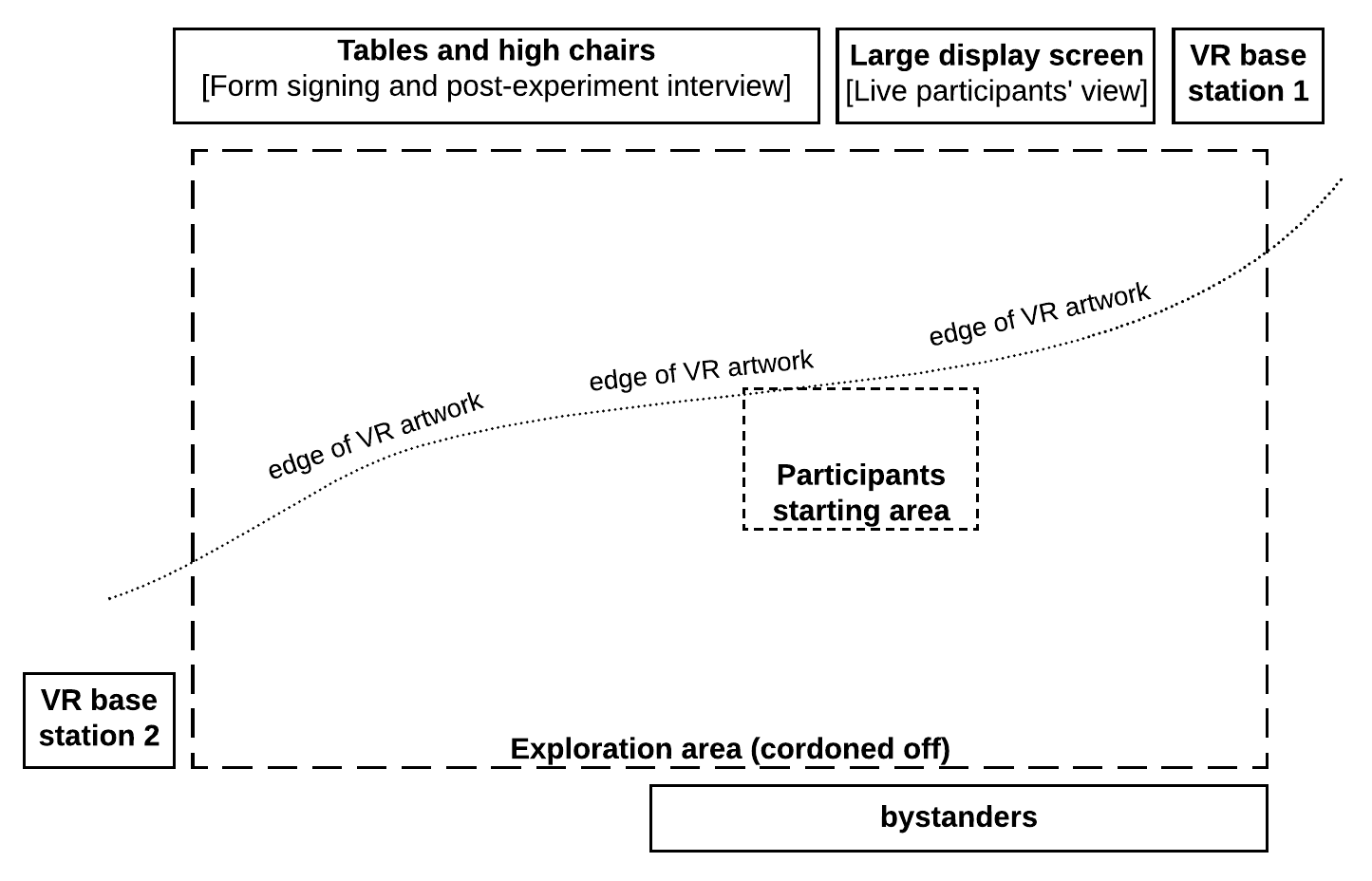}
\caption{Floor map}
\label{fig:floor}      
\end{figure}

\begin{figure}[!htbp]
\centering
  \includegraphics[trim=100 150 0 200, clip,width=0.7\textwidth]{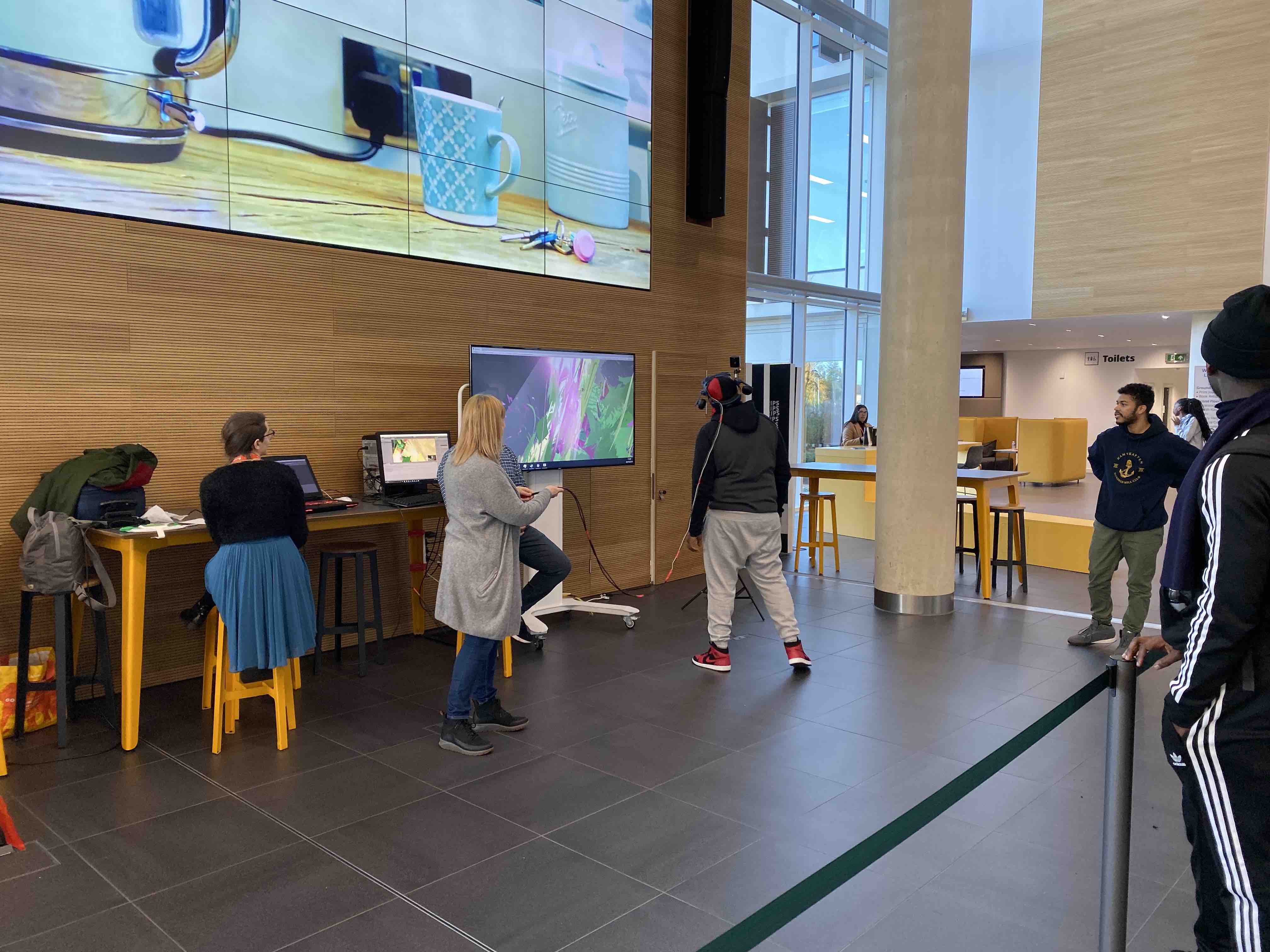}
\caption{Participants exploring the abstract VR painting}
\label{fig:user1}      
\end{figure}

The user experiment was organised in an open communal space. Figure \ref{fig:floor} and Figure \ref{fig:user1} show the arrangement of the floor space (photo taken during the experiment).

The floor area was partially cordoned off (approximately 3 by 4 meters) to enable free exploration within this space. Each participant entered the area from the right and sat by the table to fill out research consent and user information forms. Participants were asked  to answer questions on their gender, age group, game experience, knowledge of abstract art, etc. The participants were then assisted to put on a VR headset in the starting area, which is positioned near the edge of the artwork in VR. Once ready, the participants decided how to approach the artwork. An assistant helped with the cables running from the headset and ensured that the participants did not travel beyond the physical exploration area or walk into any obstacles. Any verbal comments were also recorded by the microphone on the headset. 

As the participant encountered the painting their view was displayed on a large screen. This attracted more volunteer participants and allowed the assistants to communicate with the participants about what they were seeing. A short post-viewing interview was also conducted by the artist of the VR painting in order to study how the participants felt about the VR painting encounter, how it compared to physical paintings, their opinions of interactions in VR, and the most memorable elements. 

\begin{figure}[!htbp]
\centering
    \subfigure[Gender]  {   \label{fig:gender} \includegraphics[width=0.25\columnwidth]{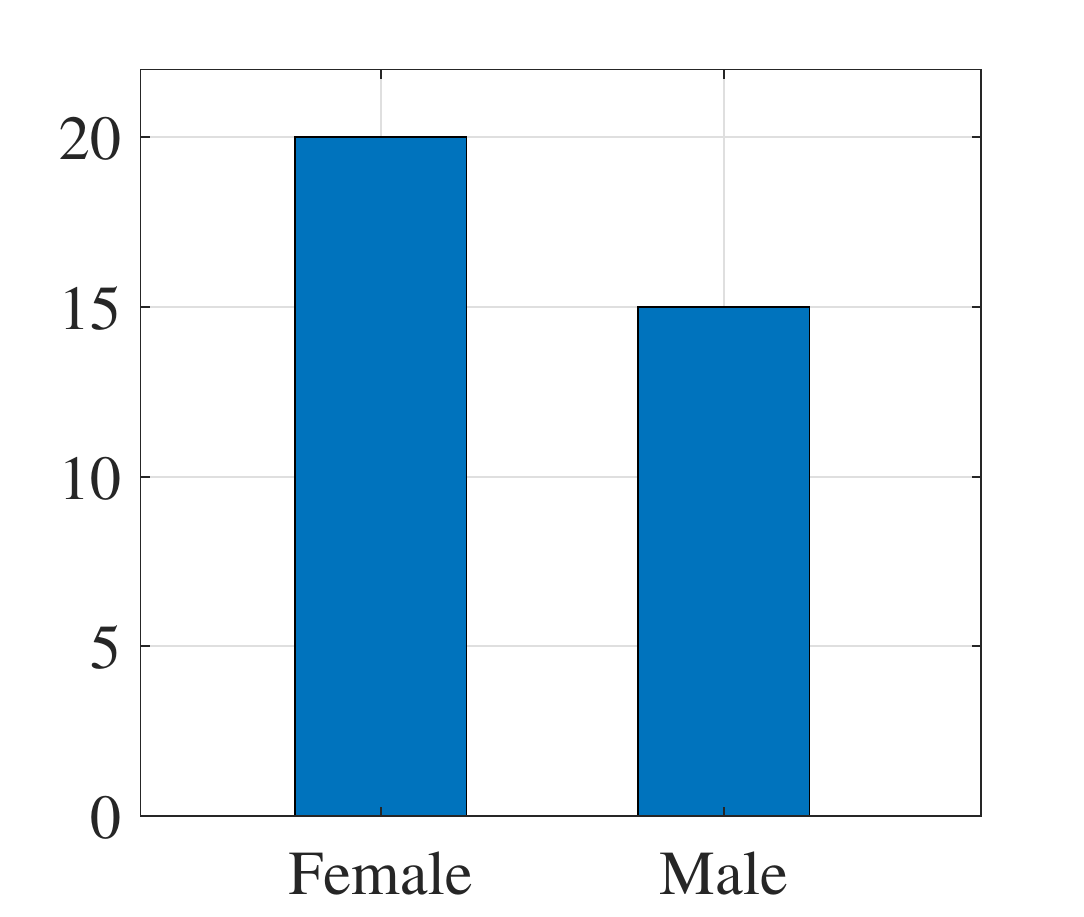} }
    \subfigure[Age group]  {   \label{fig:age} \includegraphics[width=0.25\columnwidth]{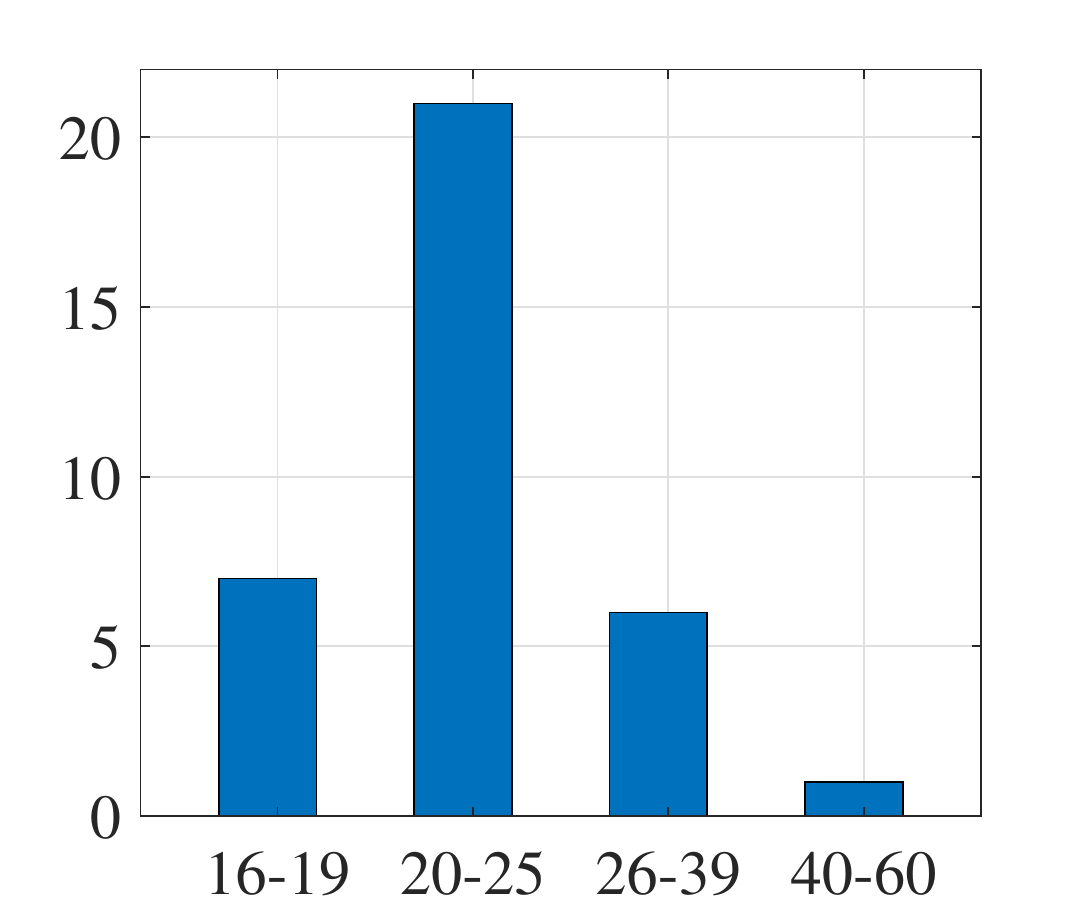} }
    \subfigure[Gaming experience]  {   \label{fig:gameexp} \includegraphics[width=0.25\columnwidth]{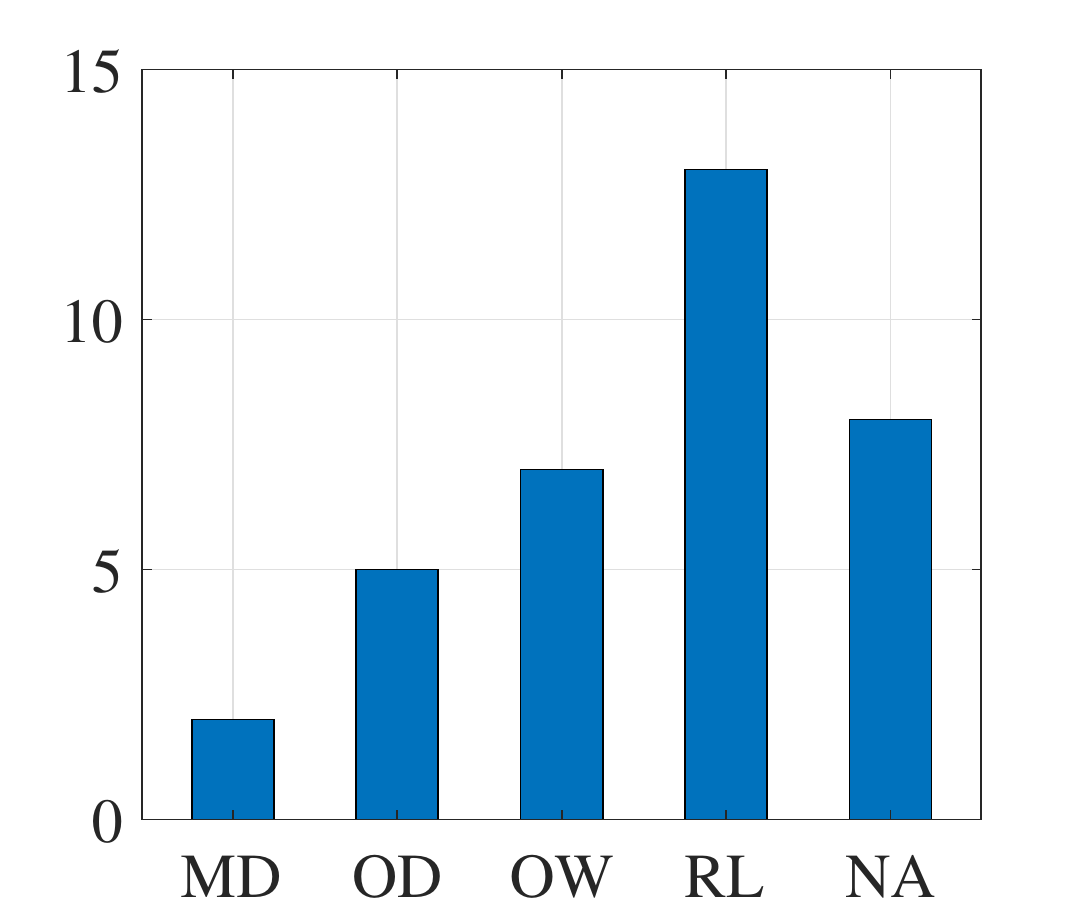}}
    \subfigure[VR experience]  {   \label{fig:vrexp} \includegraphics[width=0.25\columnwidth]{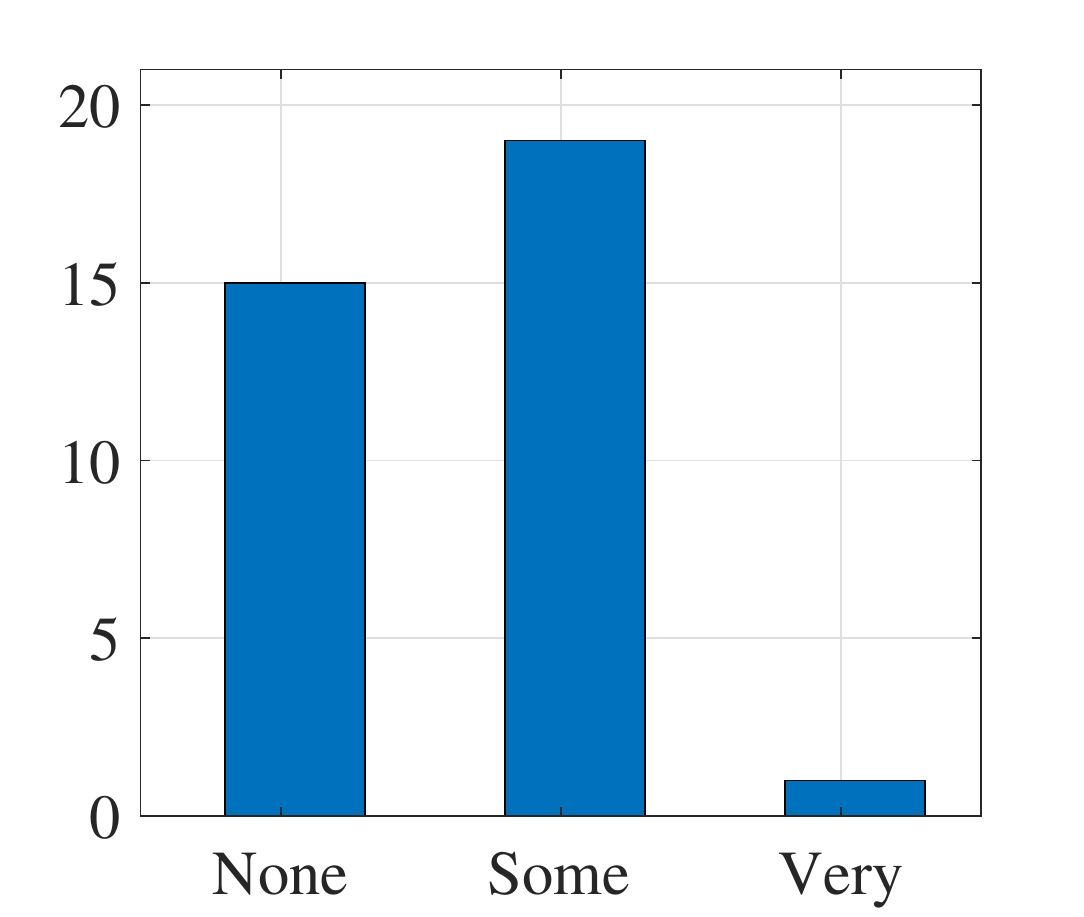}} \subfigure[Abstract art knowledge]  {   \label{fig:aaexp} \includegraphics[width=0.25\columnwidth]{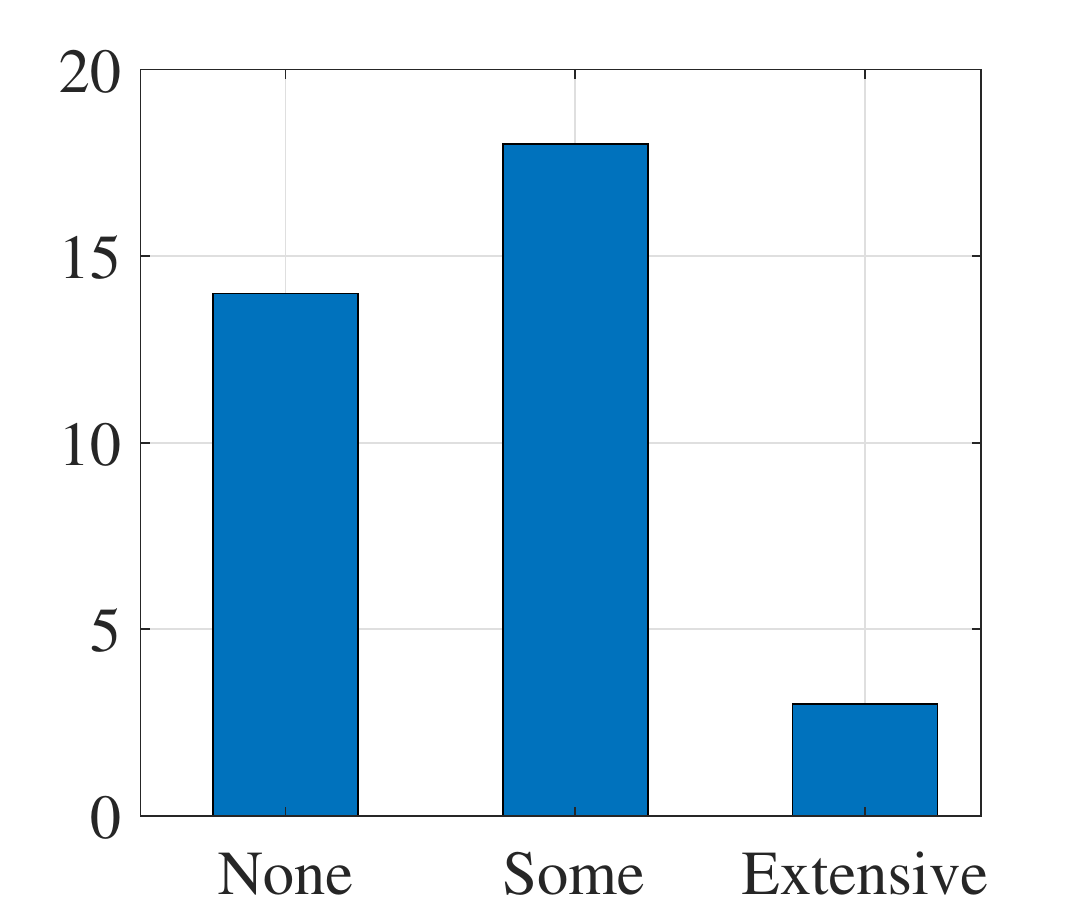} }
    \caption{Participant background}    \label{fig:userinfo}
\end{figure}

Overall, the experiment attracted 35 participants, 20 female and 15 male (Figure \ref{fig:userinfo}). The user information shows that the vast majority of the participants are aged between 16 and 25. 
More than half of the participants stated that they do not play or rarely play computer games (MD - Many times every day, OD - Once a day, OW - Once a week, RL - Rarely, NA - Not at all). Regarding their experience with VR, 15 had not tried VR before, while 18 had some experience. Only 2 participants claimed to be very experienced with VR. Similarly, only 3 participants, who studied Fine Art, had extensive knowledge of abstract painting while 18 participants were familiar with this form of artwork (Figure \ref{fig:userinfo}).

\section{Data exploration}
\label{sec:data}

\subsection{Walk}

\begin{figure}[!htbp]
\centering
  \includegraphics[width=0.9\textwidth]{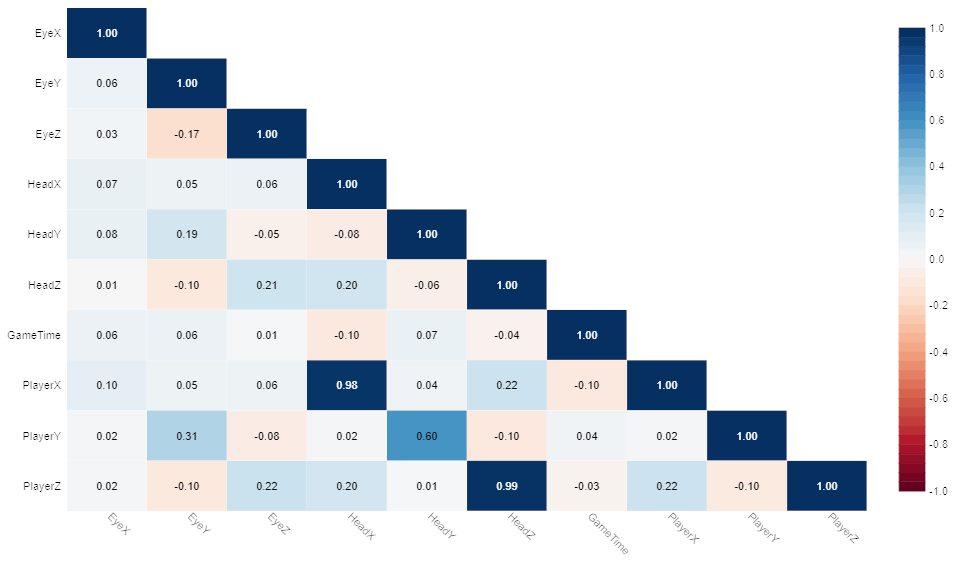}
\caption{Correlation matrix of measured data}
\label{fig:correlationmatrix}      
\end{figure}

The location of each participant was gathered by using the VR headset (head position in the physical space) and the Unity games engine (camera position in the virtual space). The location data from the two systems (head\_x, head\_y, head\_z and player\_x, player\_y, player\_z) are highly correlated (as shown in Figure \ref{fig:correlationmatrix}). Since the eye orientation data also came from the VR headset, the location data from the same source (head\_x, head\_y, head\_z) was chosen for data analysis and modelling in order to minimise asynchrony. Figure \ref{fig:correlationmatrix} also shows a very low correlation between the eye orientation and the head location as well as a low correlation between experiment time and user activity data. This indicates a low risk of collinearity in the selected inputs.

\begin{figure}[!htbp]
\centering
\foreach \i in {101,1157,1679,1719,1958,2654,2804,3003,3050,3287,3425,3703,3905,4473,4475,4646,4725,4786,5170,5982} 
{
    \subfigure[p\i]  {   \label{fig:walk\i} \includegraphics[trim=430 100 430 0,clip,width=0.17\columnwidth]{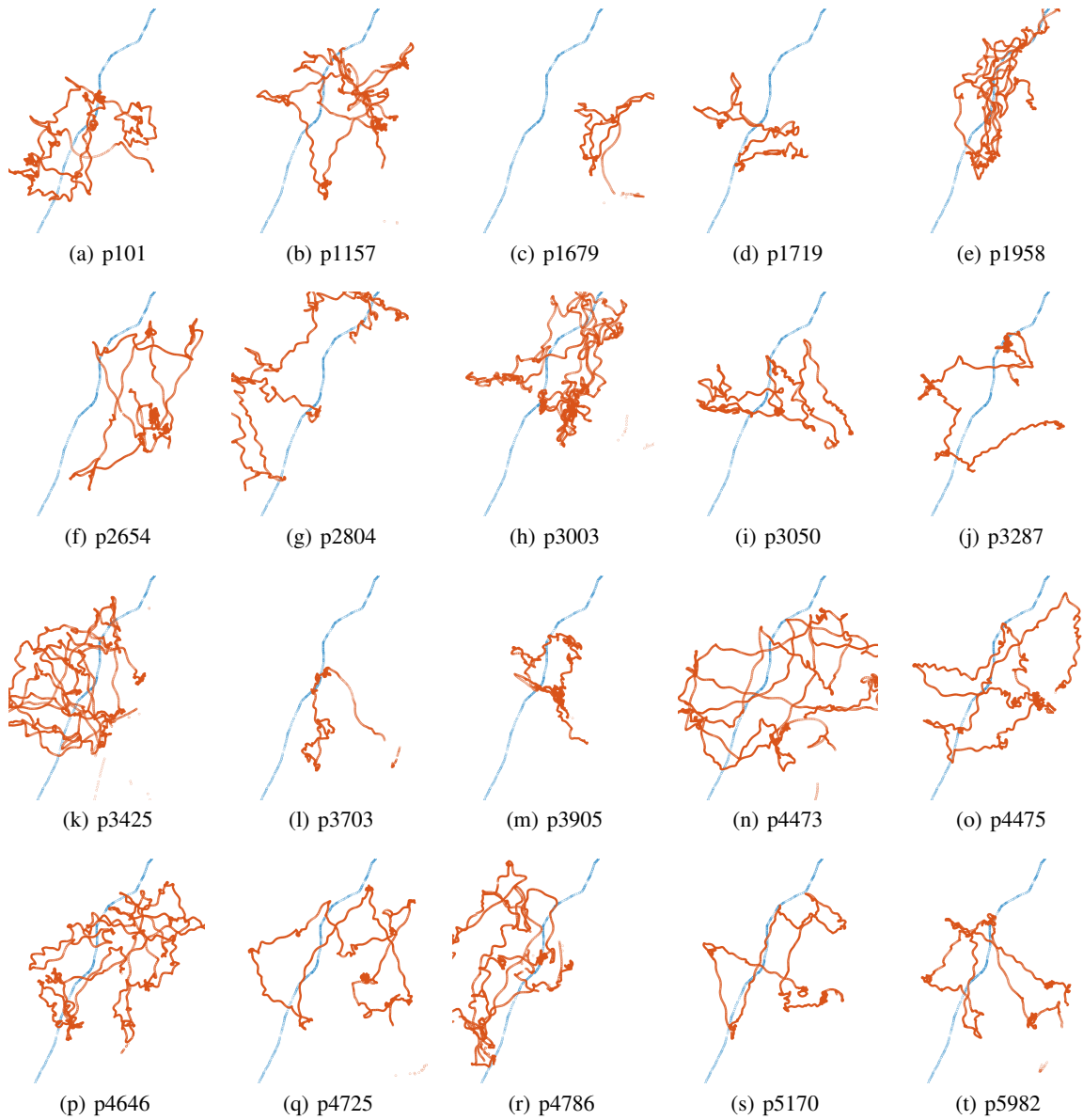} }
}
    \caption{Traces of participants' movements (top down view)}
    \label{fig:walk}
\end{figure}

Based on two coordinates \textit{head\_x} and \textit{head\_z}, Figure \ref{fig:walk} plots how 20 participants moved their locations during the experiment. Orange lines show the traces of movement and the blue lines contour the edge of the artwork at the floor level. The figure clearly demonstrates the willingness of the participants in becoming familiar with the artwork and moving to different locations to explore the artwork from different perspectives. The data also reveals completely different characteristics in the movements of participants and how the movements in VR differ from those in a physical gallery. We observed ``wanderers" such as \textit{p4473} and \textit{p4646} who travelled both inside and outside of the artwork and covered a very large area. There were ``explorers" like \textit{p1958} and \textit{p2654} who also moved extensively, but in a slightly more cautious way. Some participants like \textit{p1679} preferred to stay near the start point and mainly explored the artwork by turning their head and eyes.

\begin{figure}[!htbp]
\centering
    \subfigure[One participant]  {   \label{fig:headelevation} \includegraphics[trim=0 0 0 10,clip,width=1\columnwidth]{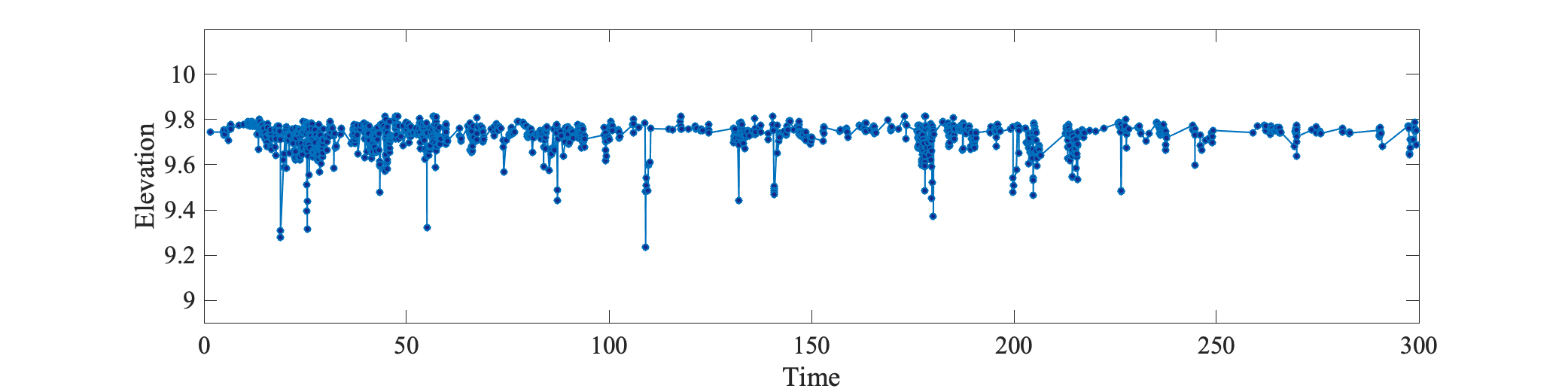} }
    \subfigure[All participants]  {   \label{fig:headelevation2}   \includegraphics[trim=110 0 120 80,clip,width=0.8\textwidth]{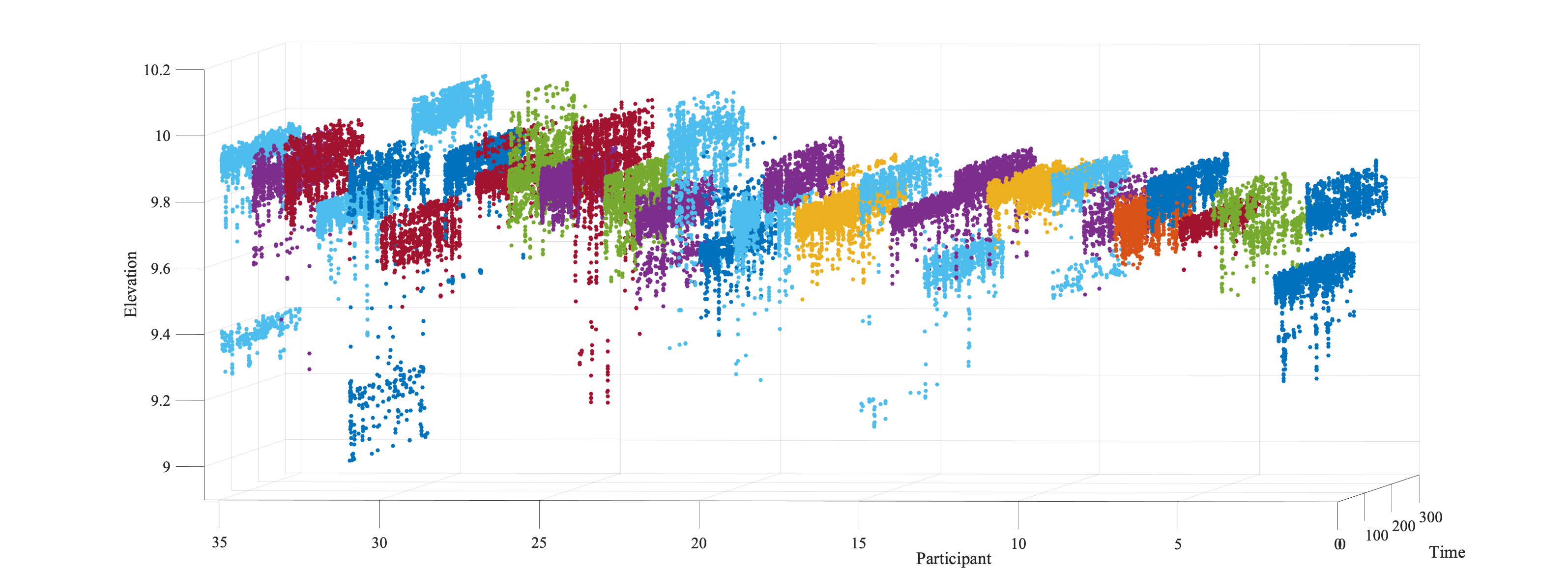}}
    \caption{Head elevation during experiment}    \label{fig:elevation}
\end{figure}

Head elevation \textit{head\_y} (Figure \ref{fig:elevation}) also shows active engagement with the artwork. Normal walking leads to small fluctuations in head elevation, while large and distinctive elevation changes indicate participants purposely lowering their heads. This seems to be to take a closer look at a much lower object (often near the ground level) or to bend over to avoid contact with a virtual ``obstacle" while exploring. Figure \ref{fig:headelevation2} shows the head elevation from all participants. Each data series has its own ``baseline'' as determined by the height of the person wearing the headset. Clearly some participants demonstrated more enthusiasm than others in getting close and interacting with the artwork. The differences in user behaviours can be attributed to different personalities and backgrounds.

\subsection{Eye orientation}

\begin{figure}[!htbp]
\centering
\foreach \i in {101,1157,1679,1719,1958,2654,2804,3003,3050,3287,3425,3703,3905,4473,4475,4646,4725,4786,5170,5982} 
{
    \subfigure[p\i]  {   \label{fig:eyes\i} \includegraphics[trim=110 20 120 90,clip,width=0.17\columnwidth]{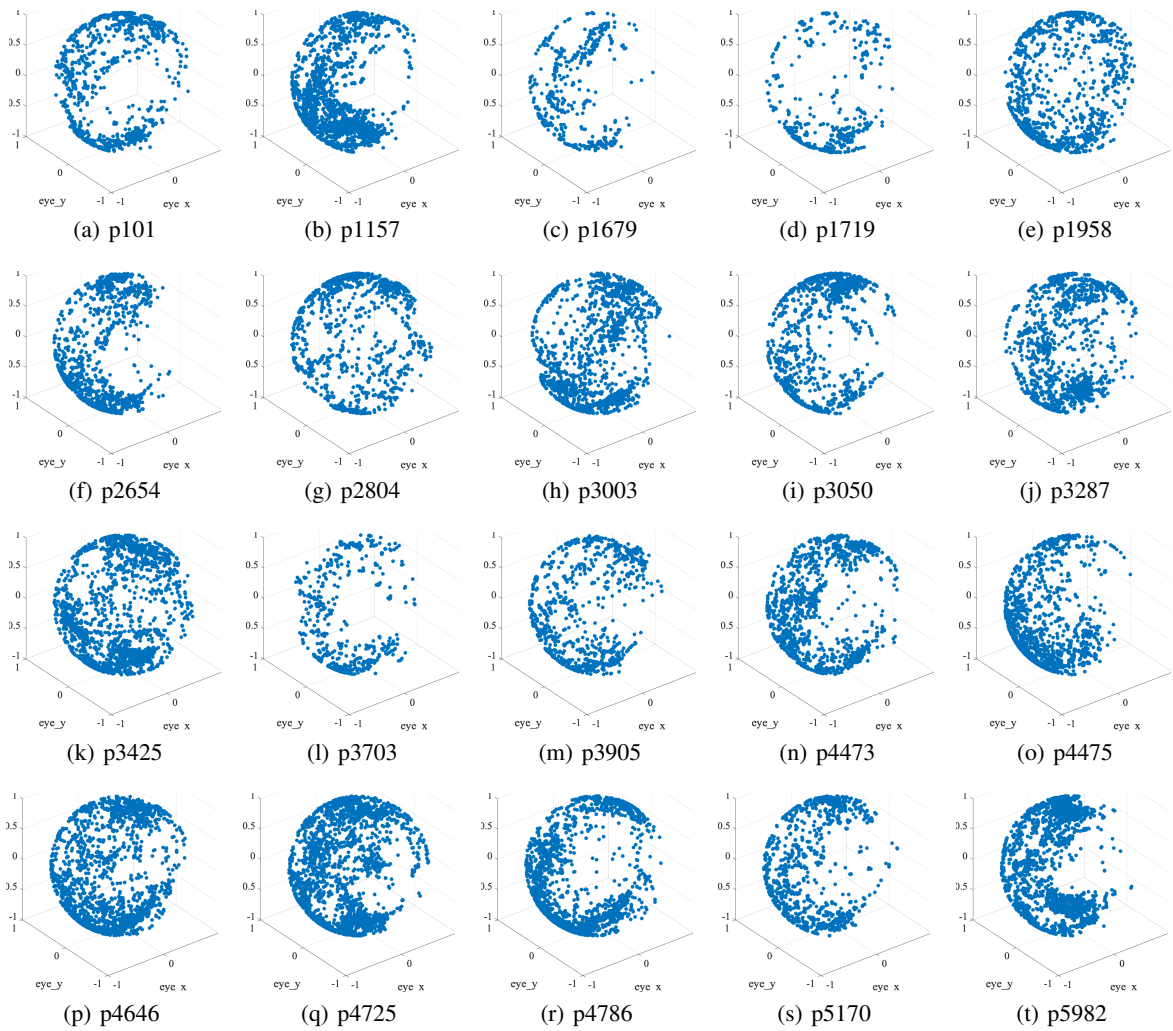} }
}
    \caption{Eye orientation}
    \label{fig:eye}
\end{figure}

As participants moved around or through the artwork, they indicated their attention on brushstrokes by how they move their heads and eyes in different directions. As human eyes are constantly moving, it is understood that there is noise and redundancy in the eye gaze data. For instance, when we move our attention from one object to a different object, our eyes may pick up many other objects in between. In the case of this experiment, there are 555,727 eye gaze records captured including groups of three types of signals: ``Focus In" which marks the moment a participant started gazing at an object, ``Normal Frame" which is a heartbeat signal registered every 30 milliseconds while the gaze on an object continues, and ``Focus Out" which flags the moment the participant moved the attention away from an object. Therefore, a one second long gaze would appear in record as one ``Focus In" followed by around 30 ``Normal Frame" then one ``Focus Out". A ``Focus out" is normally followed by another ``Focus In" when the users attention switched to a new object. Eye orientation information within each signal group is nearly identical. Hence all ``Focus In" events (a total number of 59,928) were extracted to represent all gaze events. The events captured the head location of the participant (as shown in Figure \ref{fig:walk}), the unique Brushstroke ID, as well as the three dimensional eye orientation data ({eye\_x}, {eye\_y} and {eye\_z}) as shown on a sphere in Figure \ref{fig:eye}. 

As is with the walk, the eye orientation data exhibited significantly different behaviours between the participants. In general, the participants who were more active in moving their location were more likely to have looked in different directions, as demonstrated by \textit{p3425}. Participants didn't travel deep into the artwork, for instance, \textit{p1679} often concentrated their attention on one half of the sphere as the artwork was mostly in front of them. Some dense areas at both poles of the sphere were also observed which indicates that the participants spent time investigating brushstrokes that are much lower than their eye level and also objects directly above them. No general eye orientation pattern was demonstrated across the entire participant group though some participants seem to share similar viewing preferences.

\begin{figure}[!htbp]
\centering
\foreach \i in {4092,5013,5330,3787} 
{
    \subfigure[b\i]  {   \label{fig:eyeonbrush\i} \includegraphics[trim=130 0 130 80,clip,width=0.22\columnwidth]{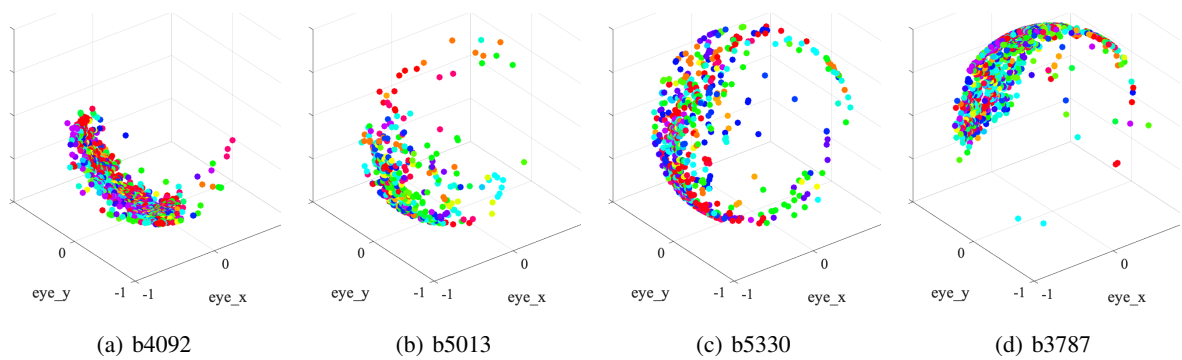} }
}
    \caption{Eye orientation on brushstrokes from participants (grouped by marker colours)}
    \label{fig:eyeonbrush}
\end{figure}

The diversity of audience behaviour is also observed at brushstroke level. Figure \ref{fig:eyeonbrush} shows the eye orientation for four different brushstrokes with data from participants separated by colour. Many brushstrokes were viewed by the participants from different angles. Individual participants also looked at the same brushstrokes multiple times from different viewpoints. We surmise that this can be attributed to a combination of 1) the inherent qualities of VR i.e., 3D artwork and free audience movements (around or through objects) which encourages more exploration and diverse viewing angles; 2) how the artist choreographs audience interactions via design choices, such as composition, colour, texture. For instance, some brushstrokes are partially visible from an angle such that participants need to change their location to discover occluded content. This is further continued where shape and lighting configuration also allow some brushstrokes to change their appearance when viewed from different angles.


\subsection{Colour encounter}

Overall the data showed that the participants encountered 64 different colours in the VR painting. To understand this further, we studied how different participants explored those colours and whether there were different preferences to colour by gender. 

\begin{figure}[!htbp]
\centering
  \includegraphics[trim=320 0 320 20,clip,width=0.95\columnwidth]{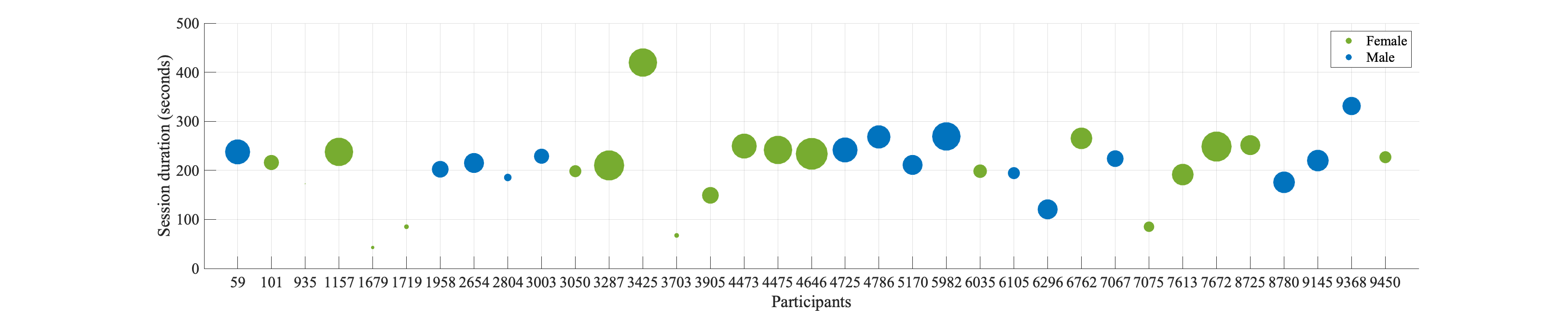}
\caption{The number of colours seen by each participant by their viewing duration and gender}
\label{fig:numberofcoloursbyduration}      
\end{figure}

Figure \ref{fig:numberofcoloursbyduration} depicts the number of colours seen by each participant, their viewing duration and gender where a larger circle indicates a higher number of colours viewed. In general it was noticeable that a longer viewing session led to more colours being encountered, with females being more likely to encounter more colours under the same condition as males.

\begin{figure}[!htbp]
\centering
    \subfigure[Female]  {   \label{fig:top15femaleduration} \includegraphics[width=0.3\columnwidth]{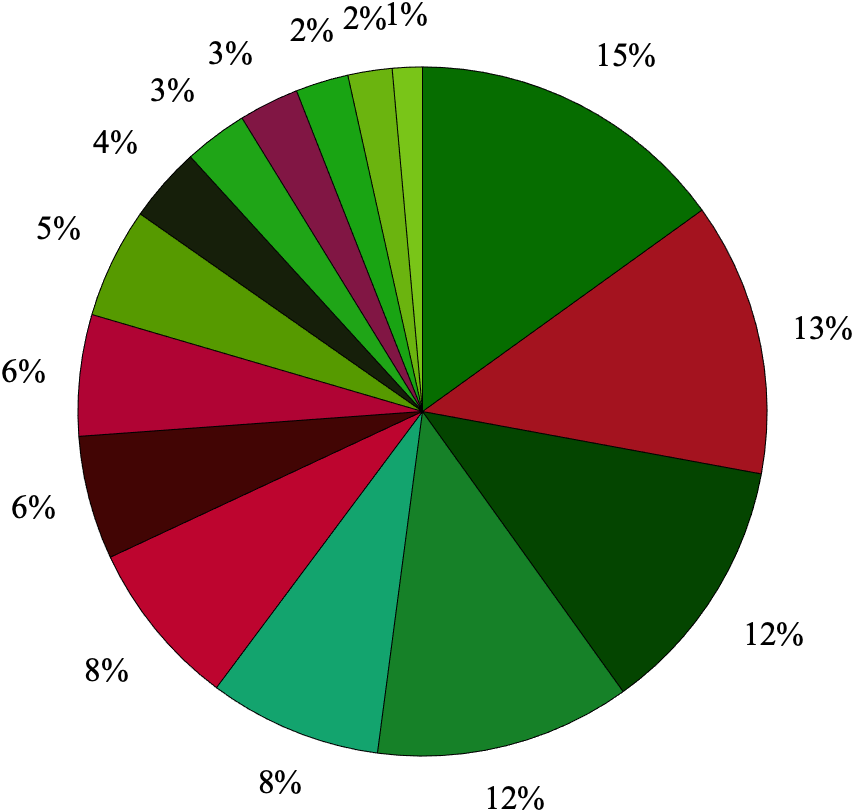} }
    \subfigure[Male]  {   \label{fig:top15maleduration}   \includegraphics[width=0.3\textwidth]{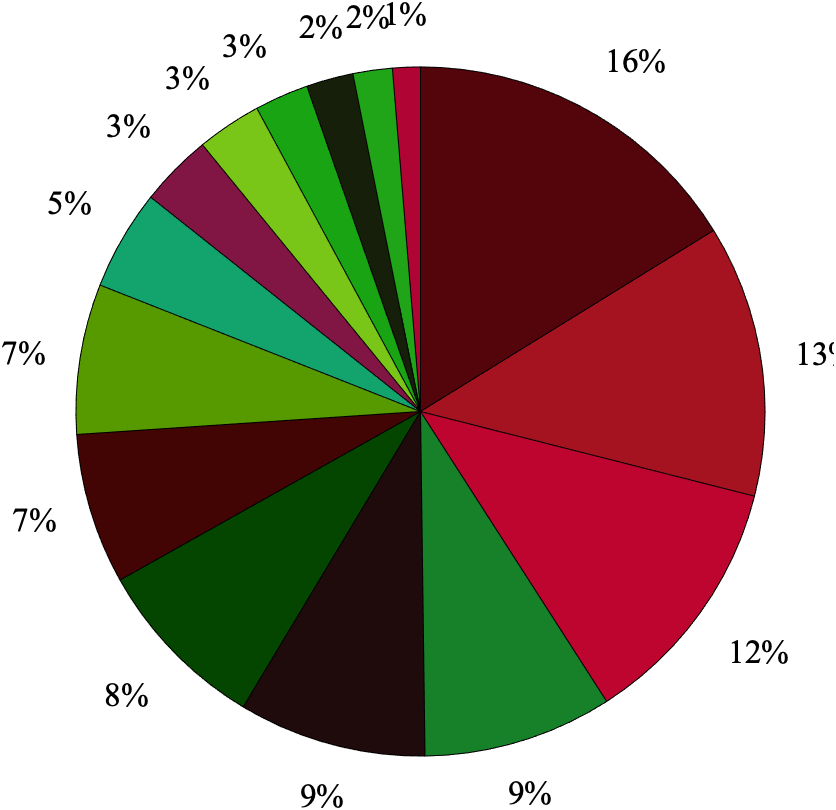}}
    \caption{Top 15 colours by gaze duration (pie charts show actual brushstroke colour)}    \label{fig:top15coloursbyduration}
\end{figure}

\begin{figure}[!htbp]
\centering
    \subfigure[Female]  {   \label{fig:top15femalecount} \includegraphics[width=0.31\columnwidth]{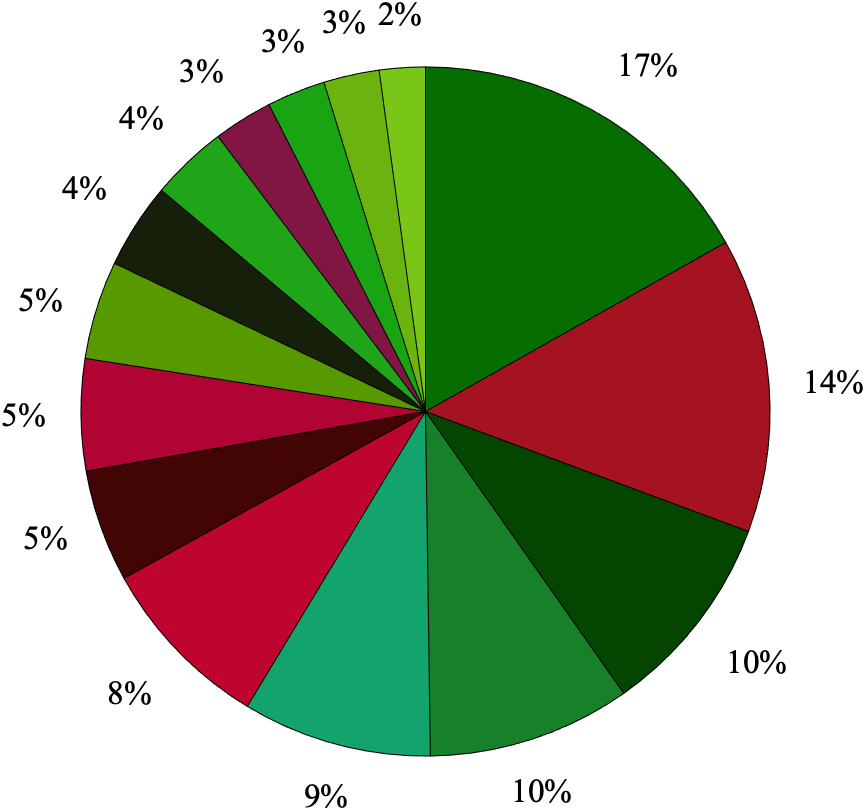} }
    \subfigure[Male]  {   \label{fig:top15malecount}   \includegraphics[width=0.3\textwidth]{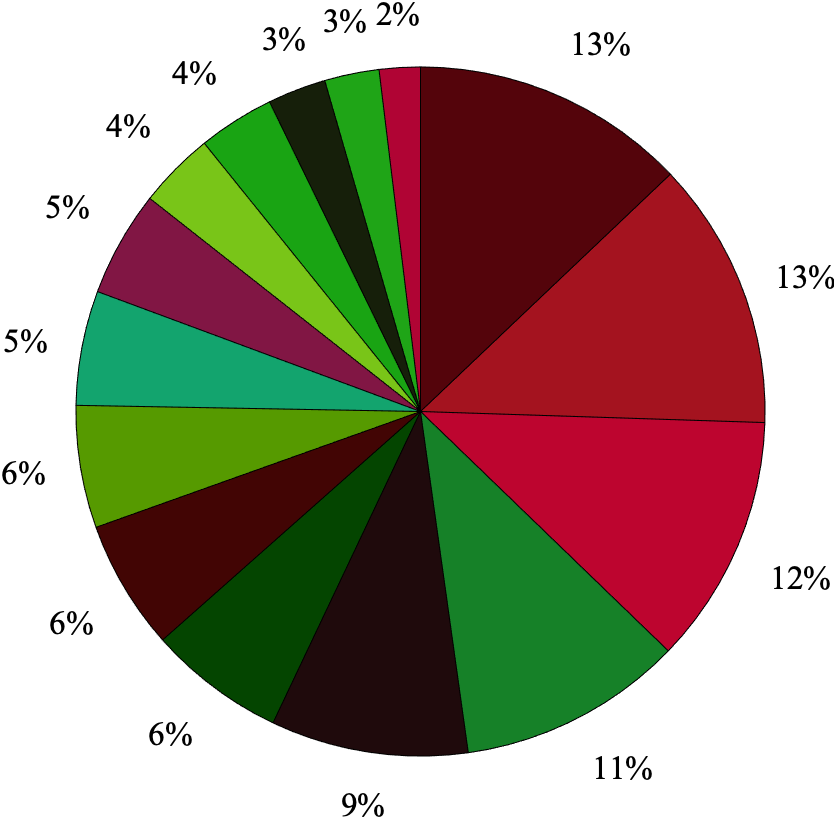}}
    \caption{Top 15 colours by gaze count (pie charts show actual brushstroke colour)}    \label{fig:top15coloursbycount}
\end{figure}

We also investigated if there was a colour preference demonstrated by gender. Figure \ref{fig:top15coloursbyduration} and Figure \ref{fig:top15coloursbycount} show the top 15 colours encountered for female and male, measured by gaze duration (how long each colour was looked at) and gaze count (how many times a colour was looked at) respectively. The pie charts' colours resemble the actual brushstroke colour in the VR painting. It is important to point out that colour is only one of the many deterministic factors for user attention. Other factors include size, shape, structure, brushstroke type, distance to the viewer and composition within the artwork. In spite of the complexity behind human attention, in this comparison between genders, it seems female participants demonstrated a preference for green, while male participants viewed red brushstrokes more in terms of both duration and count.

\subsection{Types of eye gaze}

During the experiment we observed different levels of attention demonstrated through the participants eye gaze, varying between tens of milliseconds to seconds. Gazes of different durations may correspond to different viewing habits. Some gazes are part of a quick scan while others can be a close inspection of artwork details. 

\begin{figure}[!htbp]
\centering
    \subfigure[Eye gaze clustering]  {   \label{fig:gazeclustering} \includegraphics[width=0.6\columnwidth]{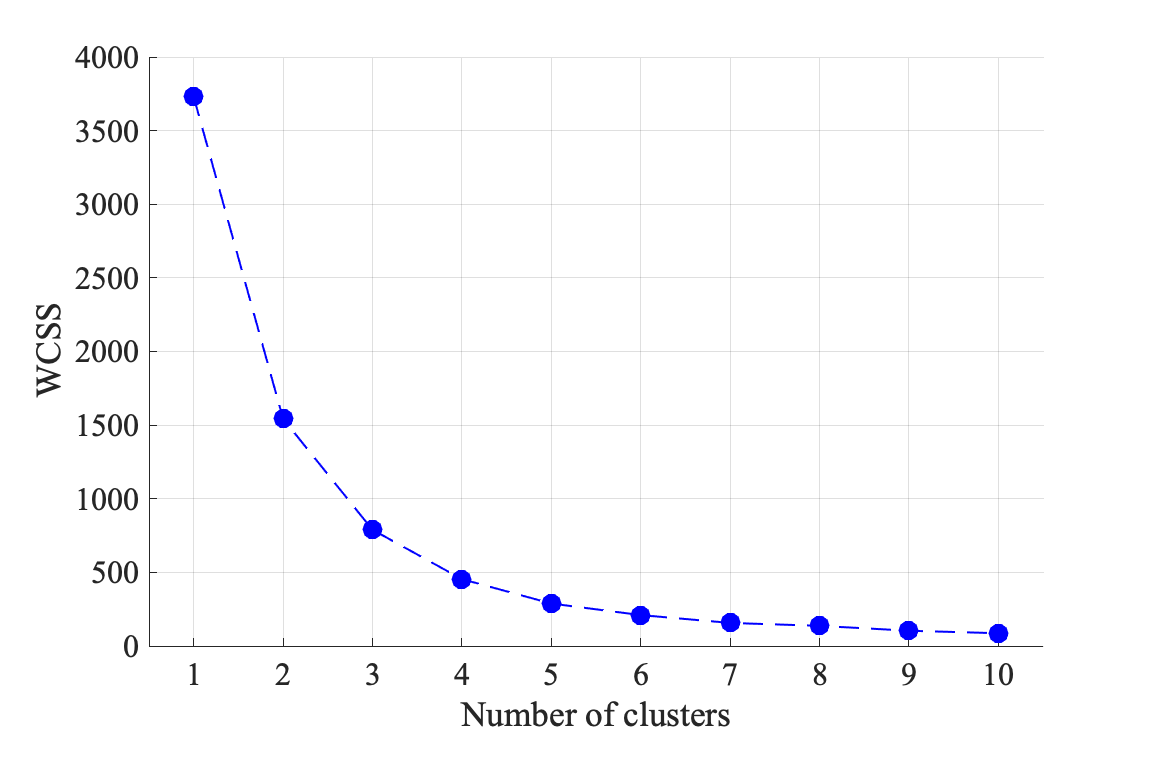} }
    \subfigure[Gaze type from participants]  {   \label{fig:gazebyuser}   \includegraphics[trim=80 0 80 0,clip,width=0.9\columnwidth]{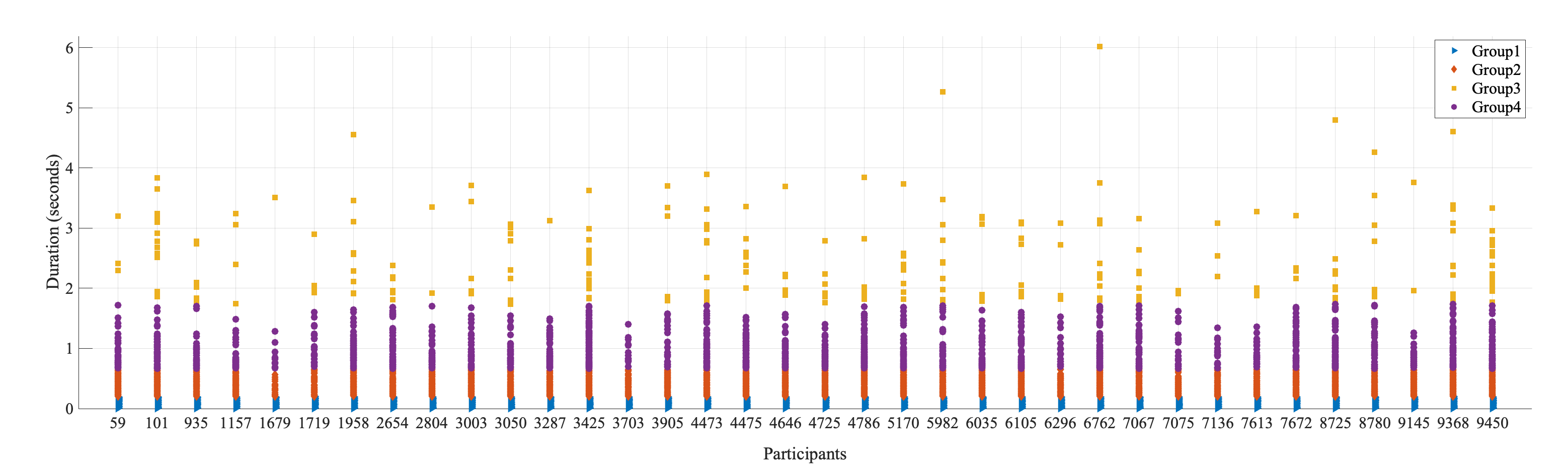}}
    \caption{User attention}    \label{fig:userattention}
\end{figure}

Using K-Means, an effective and commonly used method for clustering, the gaze data was separated into distinctive groups based on the distribution of their numerical values. Gaze events and their duration was gathered from all sessions, then K-Means method was applied with a range of configurations. Figure \ref{fig:gazeclustering} shows the Within Cluster Sum of Squares (WCSS) results when data is split into different numbers of groups. WCSS evaluates the effectiveness of clustering by measuring the squared average distance of all gaze duration within a cluster to the cluster centroid. As the number of clusters increase, the WCSS decreases, which means better clustering at the data level though too many clusters prevent a clear understanding of the levels of user attention. With four clusters, the clustering has a good balance between domain interpretability and performance. Hence, we split the level of gaze into four degrees of user attention: quick scan, normal scan, short gaze, and long gaze with the centroid of 0.047, 0.338, 0.9530, and 2.488 seconds respectively.

Participants' gaze was grouped into four clusters and plotted in Figure \ref{fig:gazebyuser}. The figure indicates that all participants demonstrated a large amount of quick scan and normal scan activities while they explored the artwork. The data clearly indicates different levels of long gaze activities across this group of participants. This shows how a piece of VR painting may elicit different experiences as participants chose their own ways to interact with the artwork. Such understandings of art encounter preferences via gaze analysis can become a main contributor for experimenting with new artwork that adapt to individual preferences.

\section{User behaviour classification}
\label{sec:modelling}

While participants' eye gaze can be clustered into groups of distinctive natures, each individual participant demonstrated a unique encounter with the experimental abstract VR painting, as was communicated by their different eye orientation and body movement. There are strong indications that some of these differences are attributed to participants' background such as gender, age, personality and related experiences/skills. If a non-intrusive quantitative measurement of human behaviour is used to accurately infer the background of the participants, it will be possible to deliver focused encounters and improved experiences for different user groups. This section provides a first attempt to classify art audiences by gender, age group, VR experience, gaming experience and art background based on six measurements: eye\_x, eye\_y, eye\_z, head\_x, head\_y, and head\_z. Table \ref{tab:datatable} shows the measurements taken from the first few seconds of a session. Questionnaire responses were used as a ground-truth labels for model training and validation.

\begin{table}[!htb]
\caption{User behavioural data}
\label{tab:datatable}
\begin{tabular}{llllll}
\hline
eye\_x     & eye\_y    & eye\_z     & head\_x   & head\_y  & head\_z  \\ \hline
-0.764599  & 0.4071314 & -0.4996187 & -12.54632 & 9.87301  & 8.309494 \\ 
-0.6116759 & 0.6332223 & -0.4741834 & -12.54806 & 9.885844 & 8.333403 \\ 
-0.6475974 & 0.6544955 & -0.3901766 & -12.51895 & 9.907124 & 8.296976 \\ 
-0.6247082 & 0.6315342 & -0.4592134 & -12.53775 & 9.901265 & 8.325699 \\ 
-0.600112  & 0.6160315 & -0.5102419 & -12.54874 & 9.898056 & 8.341338 \\ 
-0.6221098 & 0.6034535 & -0.4987954 & -12.54431 & 9.901594 & 8.338136 \\ 
-0.6538676 & 0.6154199 & -0.4401109 & -12.52308 & 9.907201 & 8.312715 \\ 
-0.5598859 & 0.6049045 & -0.5662138 & -12.53829 & 9.903415 & 8.336252 \\ 
-0.5145928 & 0.5948046 & -0.6175599 & -12.53389 & 9.904068 & 8.33251  \\ 
-0.5426044 & 0.6208919 & -0.5657312 & -12.50965 & 9.912107 & 8.294859 \\ 
-0.5358268 & 0.6092777 & -0.5844959 & -12.51862 & 9.909679 & 8.307343 \\ 
-0.5243732 & 0.6081678 & -0.5959446 & -12.5252  & 9.907698 & 8.315166 \\ 
-0.5276814 & 0.6109888 & -0.5901119 & -12.52225 & 9.908517 & 8.309989 \\ 
-0.6765658 & 0.6154901 & -0.4042251 & -12.50916 & 9.911985 & 8.287548 \\ 
-0.752409  & 0.5629492 & -0.3419628 & -12.52749 & 9.907016 & 8.312791 \\ 
-0.7641846 & 0.571066  & -0.2998287 & -12.51686 & 9.909601 & 8.296782 \\ 
-0.7675983 & 0.5718319 & -0.2894368 & -12.51374 & 9.91063  & 8.291984 \\ 
-0.7610472 & 0.565481  & -0.3178392 & -12.52108 & 9.908808 & 8.302503 \\ 
-0.7607328 & 0.5657485 & -0.3181158 & -12.52104 & 9.908838 & 8.302495 \\ 
-0.738611  & 0.551627  & -0.3874907 & -12.53943 & 9.904744 & 8.326933 \\ 
-0.7537291 & 0.5594133 & -0.3448617 & -12.52739 & 9.907848 & 8.311537 \\ 
-0.7703838 & 0.5601763 & -0.3044409 & -12.51443 & 9.911286 & 8.295473 \\ 
-0.6810743 & 0.542149  & -0.492128  & -12.54407 & 9.903443 & 8.336664 \\ \hline
\end{tabular}
\end{table}

Our session-based schema considers each of the six measurements as sequential data that independently captured the participants' behaviour during a viewing session. This means that every column in Table \ref{tab:datatable} is an independent input used to classify participants. The order of data within each sequence is retained and treated as an important feature. So visiting location A then location B is different from visiting location B first. A use case of such a session-based model would be the model observing how a person's activities change over time then predict the person's background.


The Keras\//Tensorflow framework \cite{keras,tensorflow} is used for deep learning classification. A good model performance would indicate a clear association between the tracked user behaviour and their personal background. For instance, different gender or age groups may exhibit significantly different behavioral patterns. 
Three different neural network structures were used: Feedforward Dense Network (FDN), Convolutional Neural Network (CNN) and Long Short Term Memory (LSTM) \cite{hochreiter1997long}. 

\begin{figure}[!htbp]
\centering
  \includegraphics[width=0.8\textwidth]{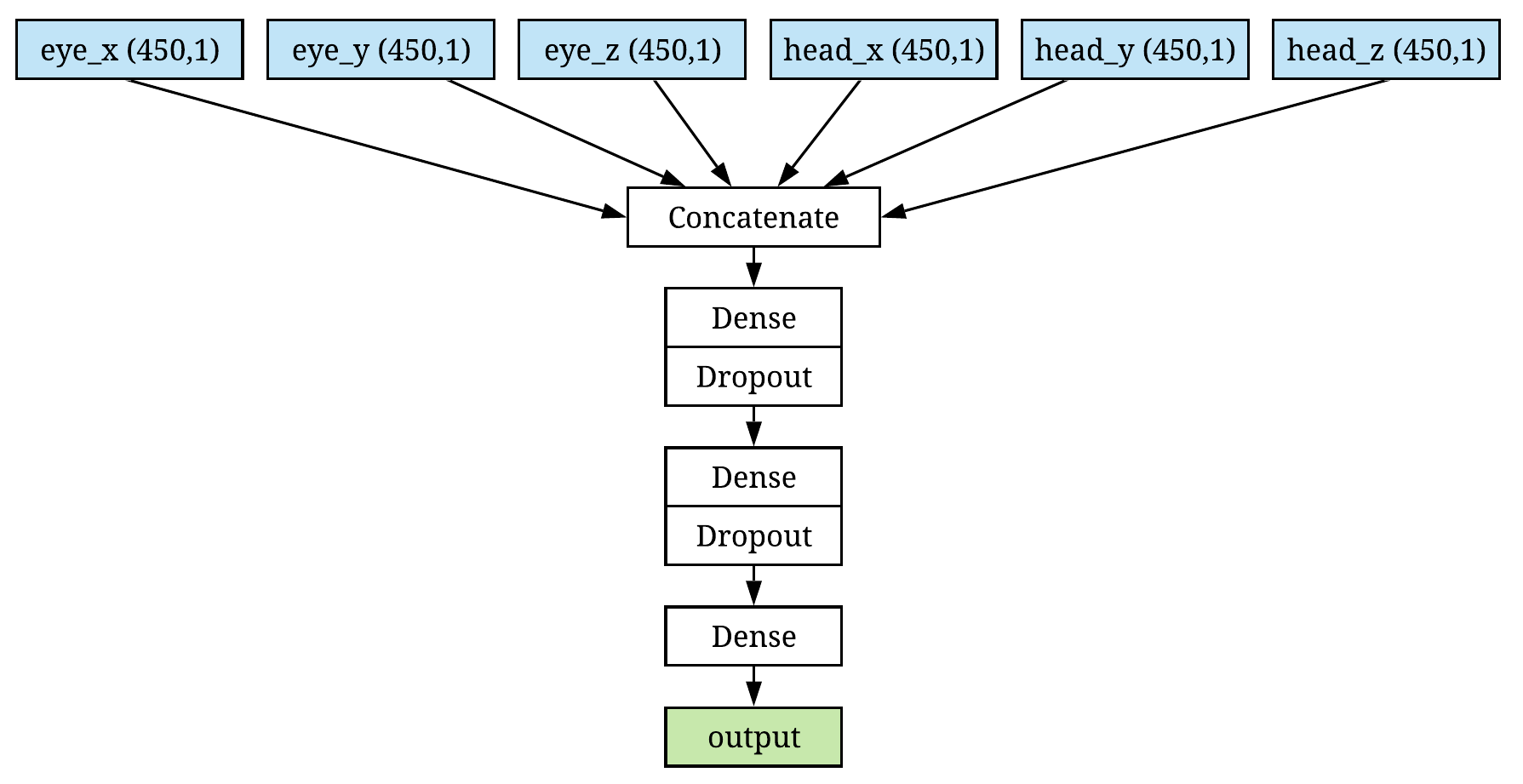}
\caption{Session-based FDN network}
\label{fig:session_dense}      
\end{figure}

\begin{figure}[!htbp]
\centering
  \includegraphics[width=0.8\textwidth]{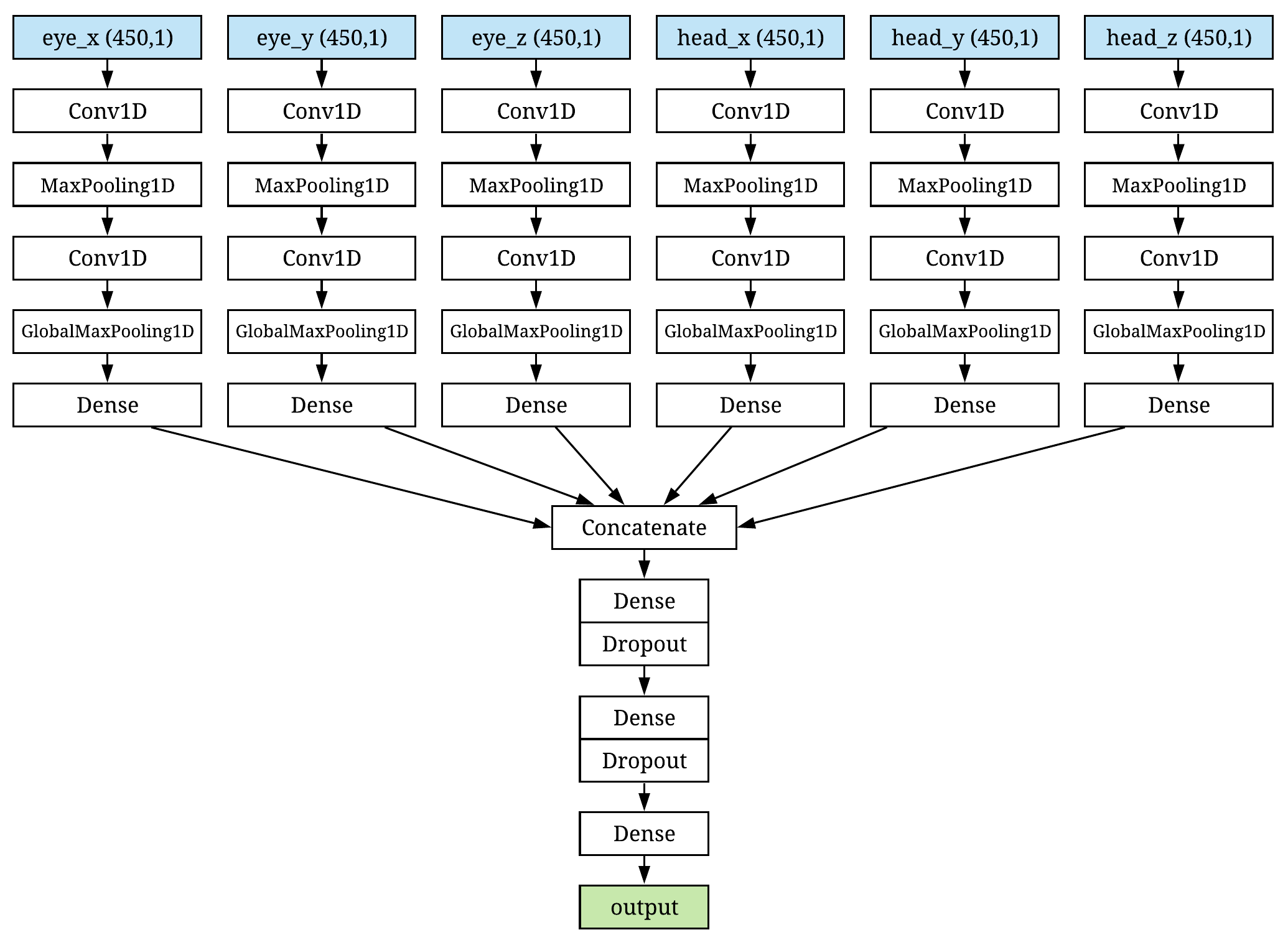}
\caption{Session-based CNN}
\label{fig:session_cnn}      
\end{figure}

\begin{figure}[!htbp]
\centering
  \includegraphics[width=0.8\textwidth]{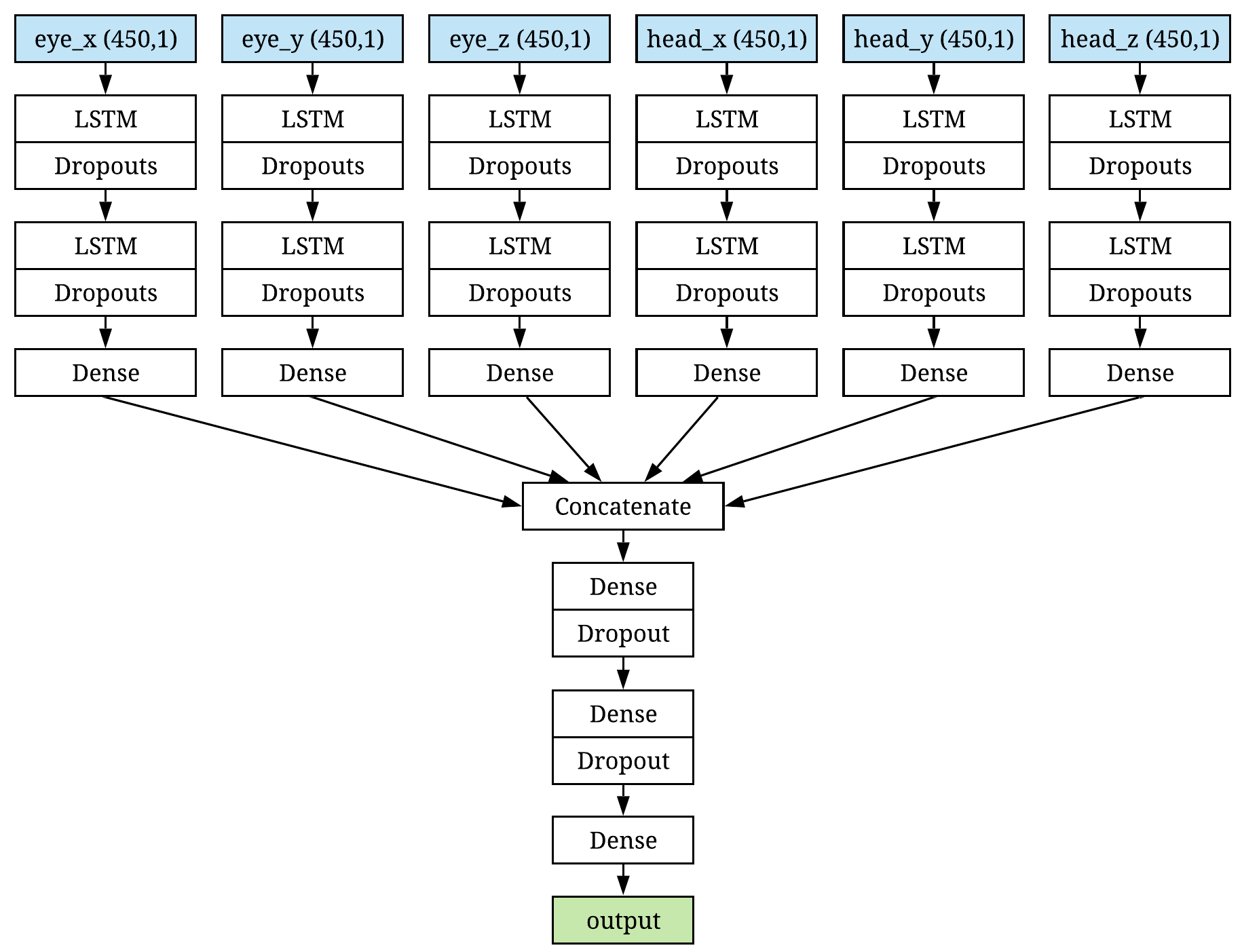}
\caption{Session-based LSTM}
\label{fig:session_lstm}      
\end{figure}

The FDN takes six inputs and concatenates them before passing the data through three Dense layers with Dropouts to mitigate over-fitting (Figure \ref{fig:session_dense}). The CNN structure adds a two-layer 1D convolutional base with MaxPooling to each of the six inputs before they are concatenated (Figure \ref{fig:session_cnn}). This should allow the network to exploit patterns in input sequential data. The patterns may reflect participants' distinctive movements such as ``leaning forward to closely inspect a paint brushstroke then moving back" and ``turning their head from left to right to see more of the painting". LSTM, as a Recurrent Neural Network (RNN) design, conducts multiple iterations of learning and uses outputs from previous steps to inform the current iteration. This allows the network to discover temporal dynamics in the sequential data. LSTM can exploit this additional ``context'' to process human behaviour. Two LSTM layers with recurrent Dropout were applied to each input layer as shown in Figure \ref{fig:session_lstm}.

As a standard practice, the output shape of each network structure adapts to each classification task. For gender prediction, the output shape is set to \(1\) (binary) via a \textit{sigmoid} activation function at the last layer, as all participants chose either female or male in the pre-test questionnaire. Other predictions have categorical outputs via a \textit{softmax} activation function and the shapes are configured to match the number of label categories. 

Since the dataset contains numeric values of different data ranges and scales, the data is normalised for each of the six measurements. Because a strict rule was not placed on how long a viewing session should be, some participants spent more time than others. For session-based models, the input data from all 35 participants must be in the same shape. Therefore, the first 450 Focus-In events when participants started gazing a brushstroke were sampled from each participant for the modelling. The deep learning training on each of the three network structures was carried out 20 times independently. Each run started with a random 70/30 split on the dataset so that 70\% of the data was randomly selected for training while the rest were used for validation. It was made sure that the data from any participant will fall in either the training set or the validation set. This ensured that the participants in the training set were entirely unknown to the validation process. A maximum of 200 epochs for each run was set but Keras' ``EarlyStopping" callbacks was used to stop the training process and restore the best weights when the validation loss had stopped decreasing in several epochs. The training was carried out on an HP Workstation with a Nvidia TITAN RTX graphics card for GPU acceleration.

\begin{figure}[!htbp]
\centering
    \subfigure[FDN]  {   \label{fig:session-dense} \includegraphics[width=0.4\columnwidth]{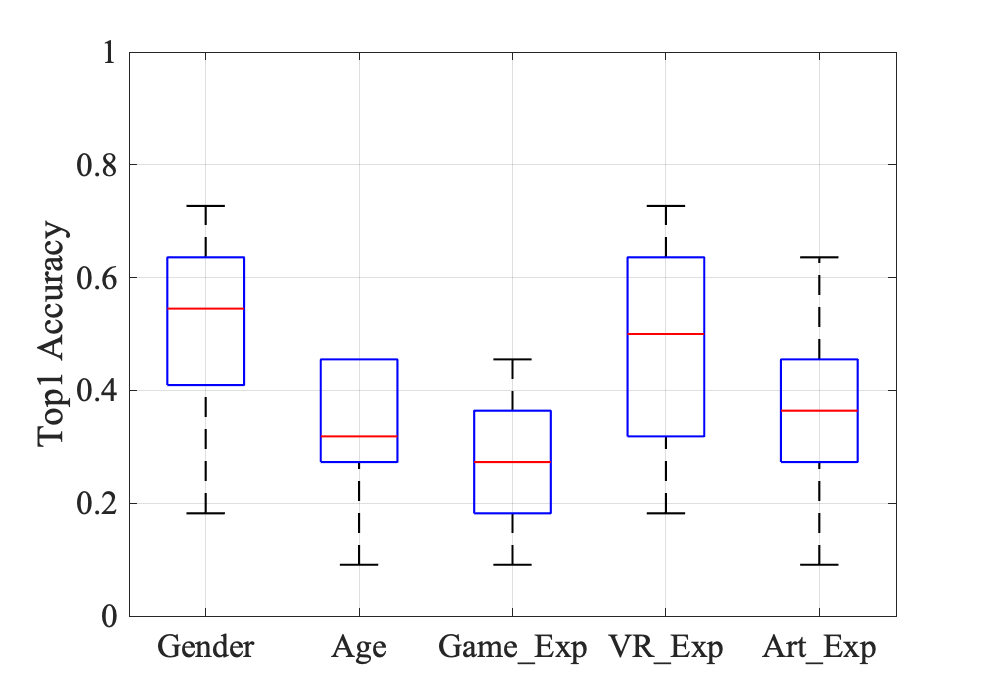} }
    \subfigure[CNN]  {   \label{fig:session-cnn}   \includegraphics[width=0.4\textwidth]{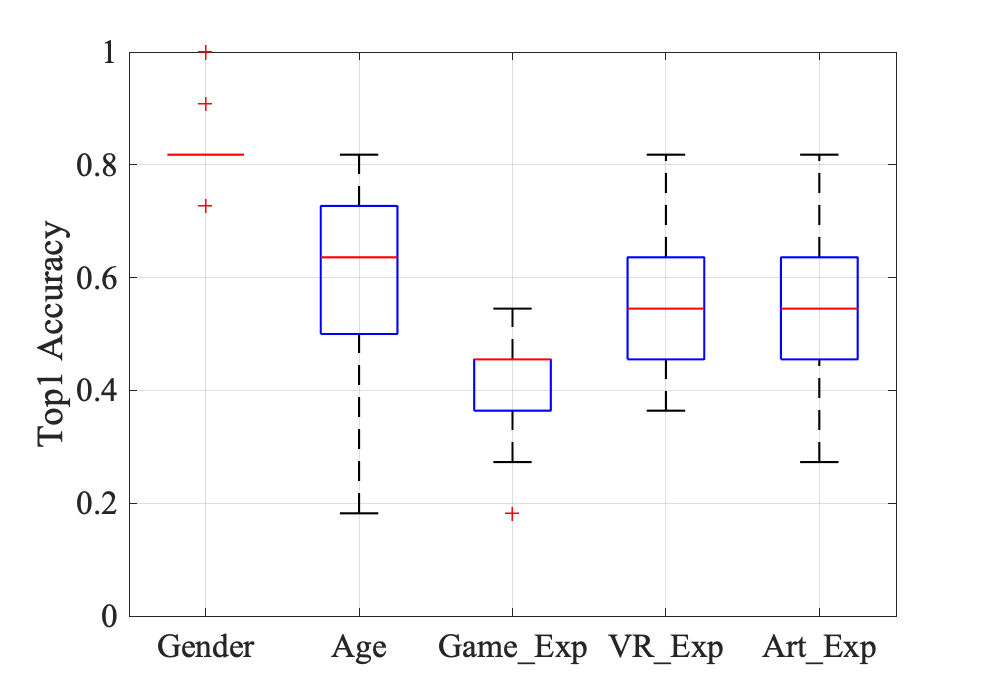}}
    \subfigure[LSTM]  {   \label{fig:session-lstm}   \includegraphics[width=0.4\textwidth]{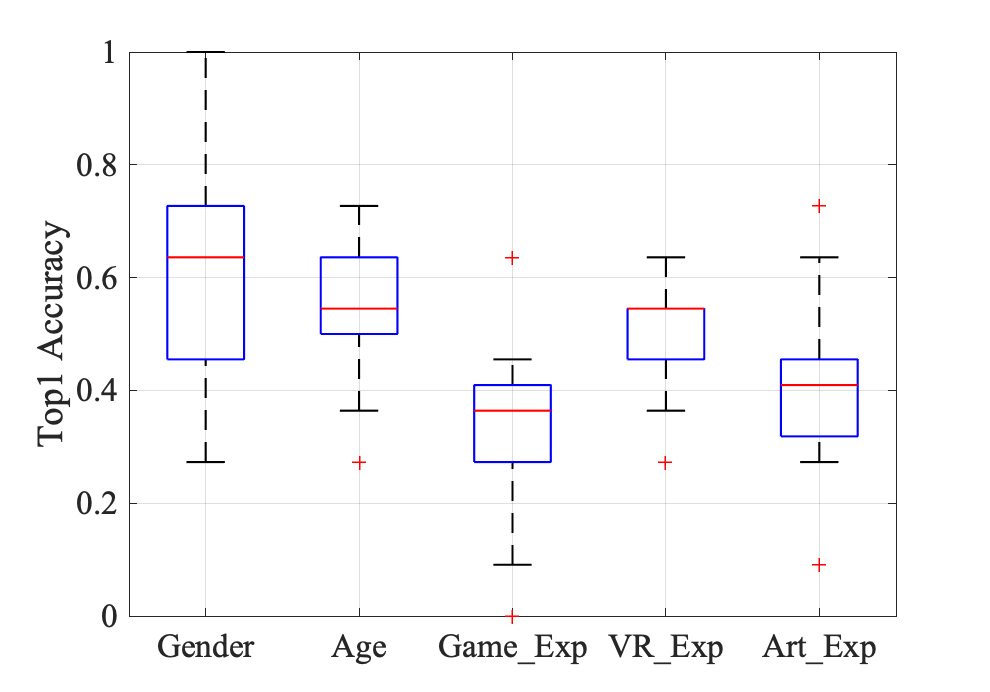}}
    \caption{Performance of session-based models (Top1 accuracy)}    \label{fig:session-based-modelling-results}
\end{figure}

Figure \ref{fig:session-based-modelling-results} shows the boxplots (from 20 runs) of the Top1 accuracy of our models predicting gender, age group, game experience, VR experience, and abstract art knowledge. The CNN model exhibits the best performance in all predictions. The median of the results given by the CNN model is 0.82, 0.64, 0.46, 0.55 and 0.55. For a dataset of 35 participants, the results are excellent for gender prediction and good for age prediction. This indicates the cross-gender difference in patterns of head and eye movements when participants interact with VR artwork and how the patterns can be used for classification. Age was more difficult to predict using behavioural patterns though its accuracy is significantly higher than a random prediction (with an accuracy of 0.25). Meanwhile, it seems that participants with different levels of game experience, VR experience, and abstract art knowledge did not show significantly different behavioural patterns as they moved between different parts of the artwork. 

It is believed that increasing the number of VR participants would improve model performance and support the training of more complex neural network structures to potentially discover deeper behavioural patterns for background classification. Furthermore, classification based on subjective opinions (e.g., self-assessment of game experience, VR experience, and abstract art knowledge) can be more challenging than more objective factors such as gender and age group, so further work would also be required to clarify this area.

\begin{figure}[!htbp]
\centering
    \subfigure[FDN]  {   \label{fig:session-dense} \includegraphics[width=0.45\columnwidth]{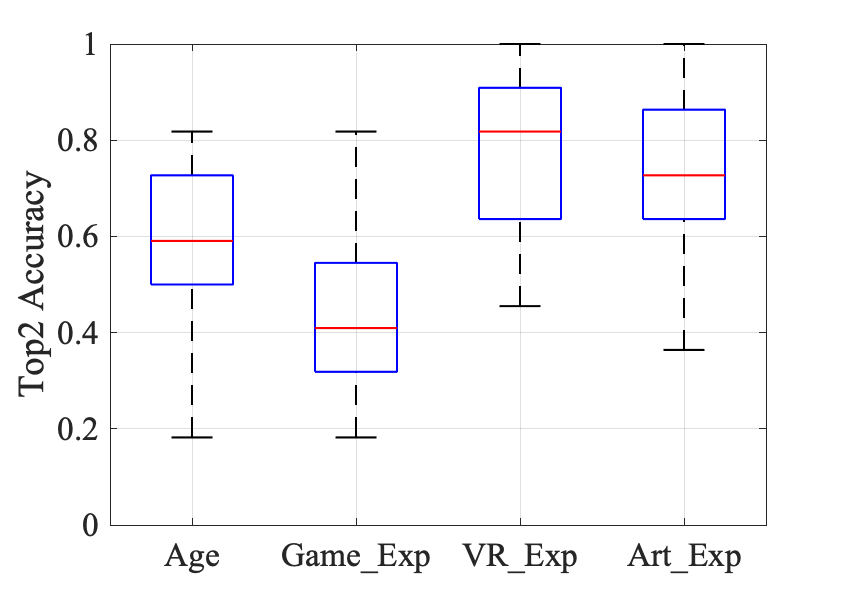} }
    \subfigure[CNN]  {   \label{fig:session-cnn}   \includegraphics[width=0.45\textwidth]{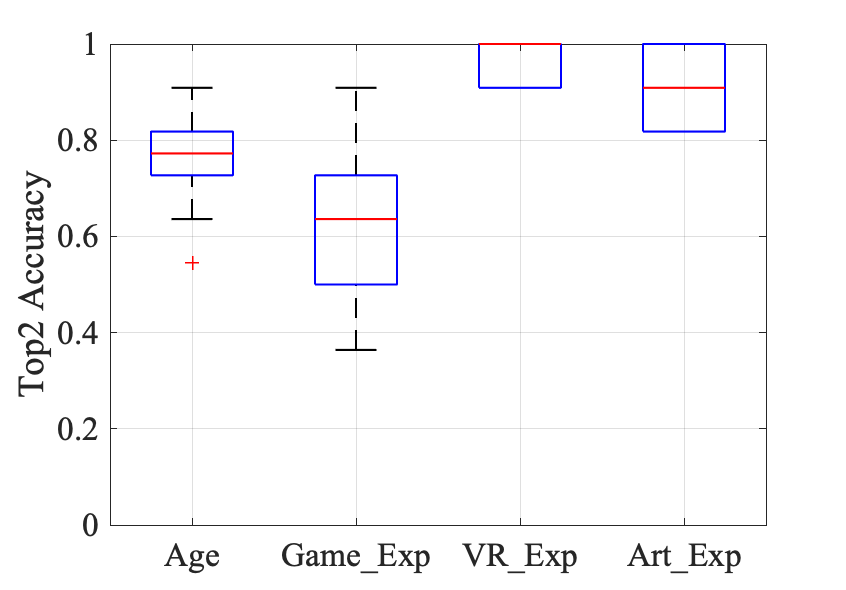}}
    \subfigure[LSTM]  {   \label{fig:session-lstm}   \includegraphics[width=0.45\textwidth]{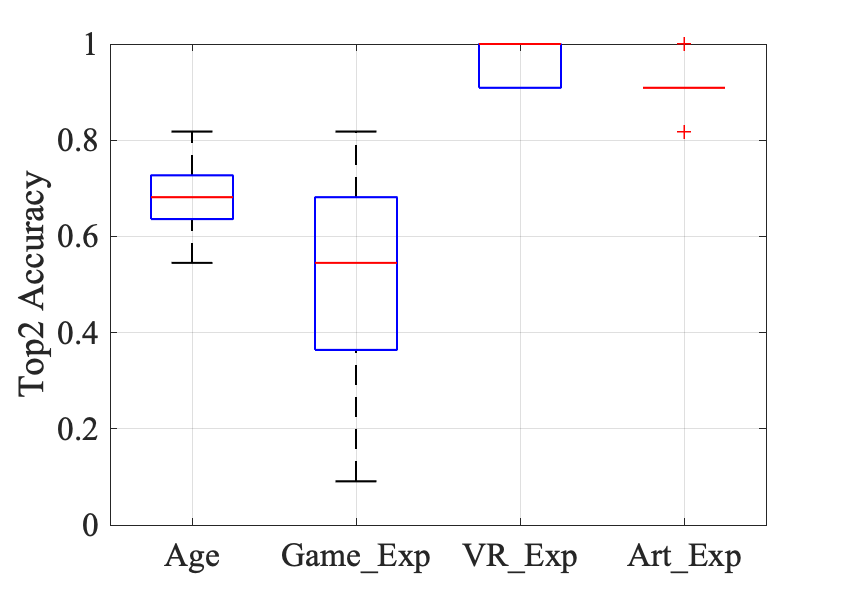}}
    \caption{Performance of session-based models (Top2 accuracy)}    \label{fig:session-based-modelling-results-top2}
\end{figure}

The Top2 accuracy of four categorical classifiers was also studied (Figure \ref{fig:session-based-modelling-results-top2}). Some predictions clearly showed significant performance above 0.8. Although this is mainly attributed to the low number of categories, the Top2 accuracy is a positive indicator of how the models capture any possible relationships between human background and behaviour.





\section{User attention visualisation}
\label{sec:visualisation}

\subsection{VR heat map}

One of the main objectives of this work is to develop a tool to visualise user attention on VR artwork as a feedback channel for VR content creators. A common method to visualise human attention is a 2D heat map of gaze intensity superimposed on the original content. The heat map is often semi-transparent so that the original objects remain partially visible. For VR, a 2D heat map can be added as an augmentation layer between a viewer and the artwork. Besides the negative impact to immersion, it becomes problematic when associating a 2D mark to a 3D brushstroke, especially in densely populated areas where brushstrokes intertwine. As such, the 2D heat map would also have to adapt to different viewing angles and distances. If an artwork is reviewed by multiple viewers, then each viewer would require a separate heat map instance. Hence we suggest a 2D heat map would be more suited to static viewing on a conventional monitor and not fit for data visualisations in VR. 

\begin{figure}[!htbp]
\centering
    \subfigure[Opacity-based attention visualisation]  {   \label{fig:opacity} \includegraphics[width=0.65\columnwidth]{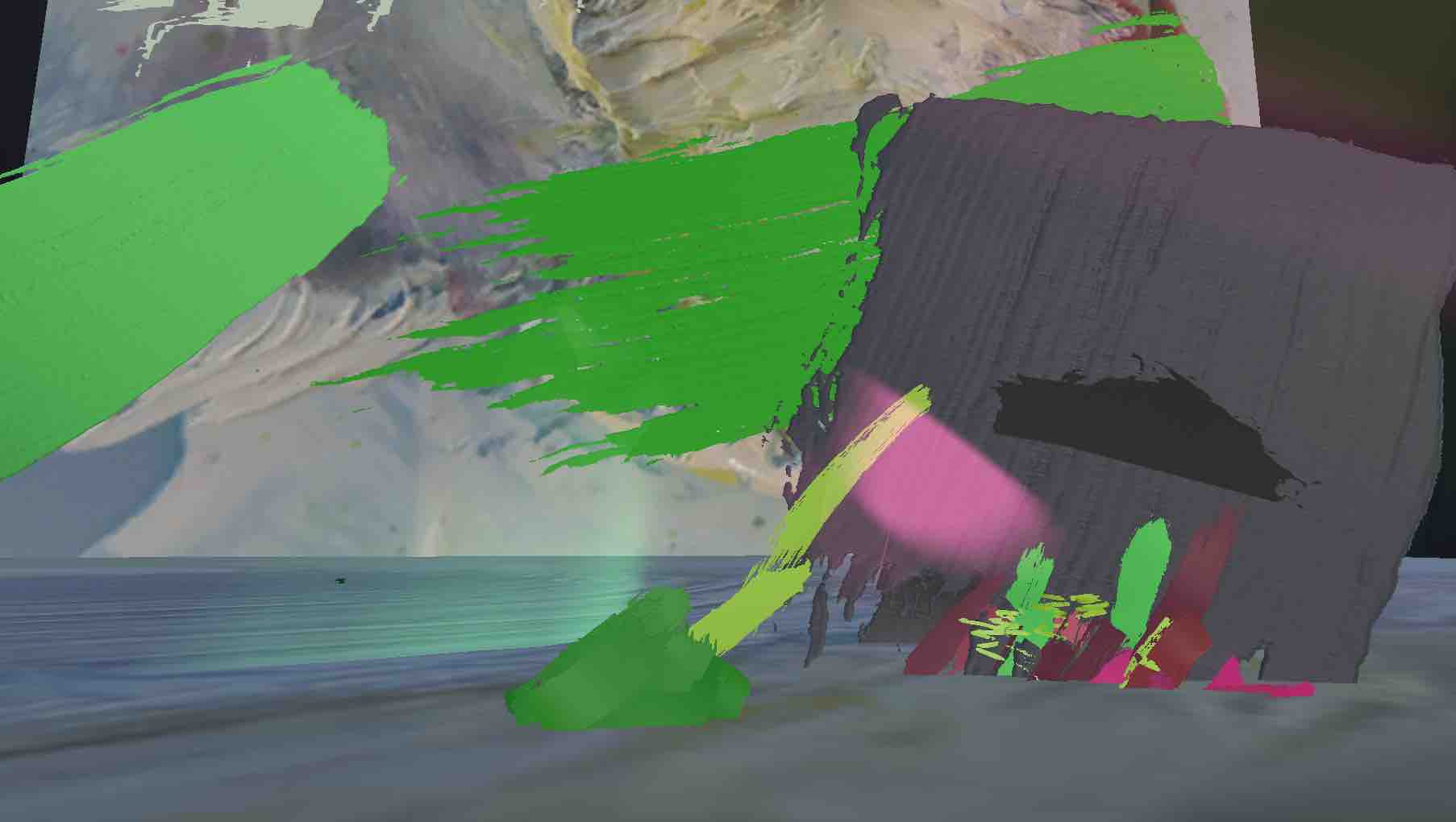} }
    \subfigure[Saturation-based attention visualisation]  {  \label{fig:saturation}
    \includegraphics[width=0.65\columnwidth]{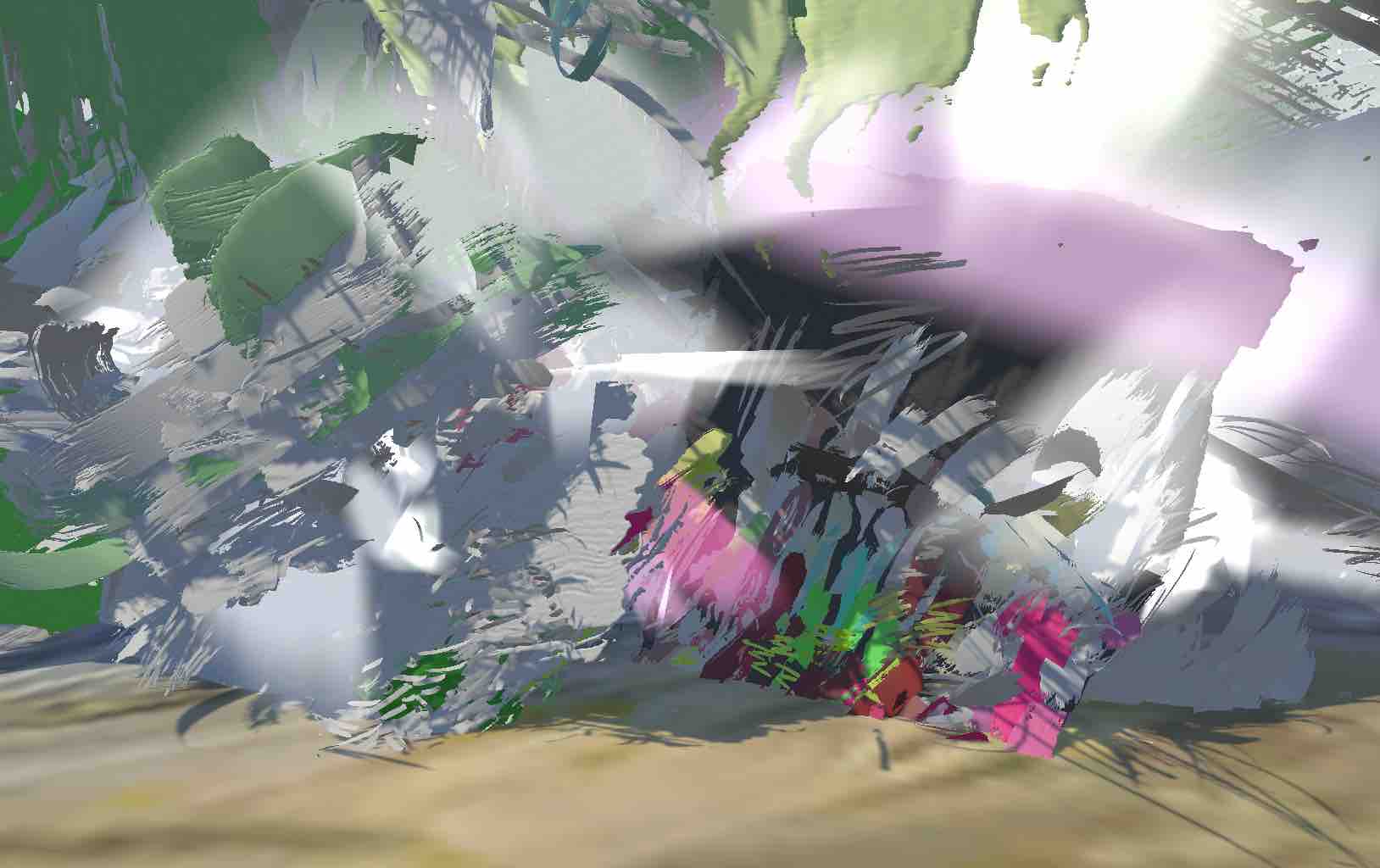} }
    \caption{User attention visualisation}    \label{fig:user-attention}
\end{figure}

Recent advances in information visualisation within games design are used to improve the gaming experience. For instance, player health and ammunition levels are not displayed as an overlay at a corner of the screen but integrated as part of the game character or an item of equipment that they carry. In alignment with this idea from games design, this work experimented with two methods of visualising the viewers' attention in VR by controlling the opacity and colour saturation of the brushstrokes based on the level of attention received from 35 participants. The implementation was done by a customised script based on Tilt Brush Unity Toolkit \cite{tiltbrushtoolkit}.

Figure \ref{fig:opacity} shows the results when there is a single opacity threshold applied to the artwork by altering its alpha channel. In this particular example, the opacity of a brushstroke is set to 1 if it has received more than 1 second of attention per viewing session (35 seconds in total for the experiment). Otherwise, the opacity is set to 0. This screening method clearly demonstrates which parts of the artwork viewers were mostly interested in on average, but the removal of other parts makes it difficult to analyse the users' interest in its context. In addition, finer control of opacity was also tested. However most Tilt Brush brushes are natively opaque and cannot be set semi-transparent directly.

For the colour saturation-based visualisation, the R,G,B values of brushstrokes are altered (Figure \ref{fig:saturation}). Any brushstroke that received user attention above a threshold keeps its original colour. Other brushstrokes are de-saturated based on how far their received attention is from the threshold. The process factored in, the original brightness of brushstrokes and human visual system's response to colours, in order to preserve the aesthetic of the artwork as much as possible. This means a bright colour will retain brightness after being de-saturated or converted to a neutral colour. The saturation-based visualisation preserves the structure of the artwork, whilst revealing how user attention moves from one ``hotspot" to another. 


\subsection{Verbal response}

The verbal responses during the VR encounter can also provide valuable feedback to the content creators. The headset's microphone turned out to be an ideal choice to capture participants' voice with low background noise due to its close proximity to the source. Since there was no instructions for the participants to provide in-test feedback, the recordings reflect their genuine feelings and emotions.

\begin{figure}[!htbp]
\centering
  \includegraphics[width=0.8\textwidth]{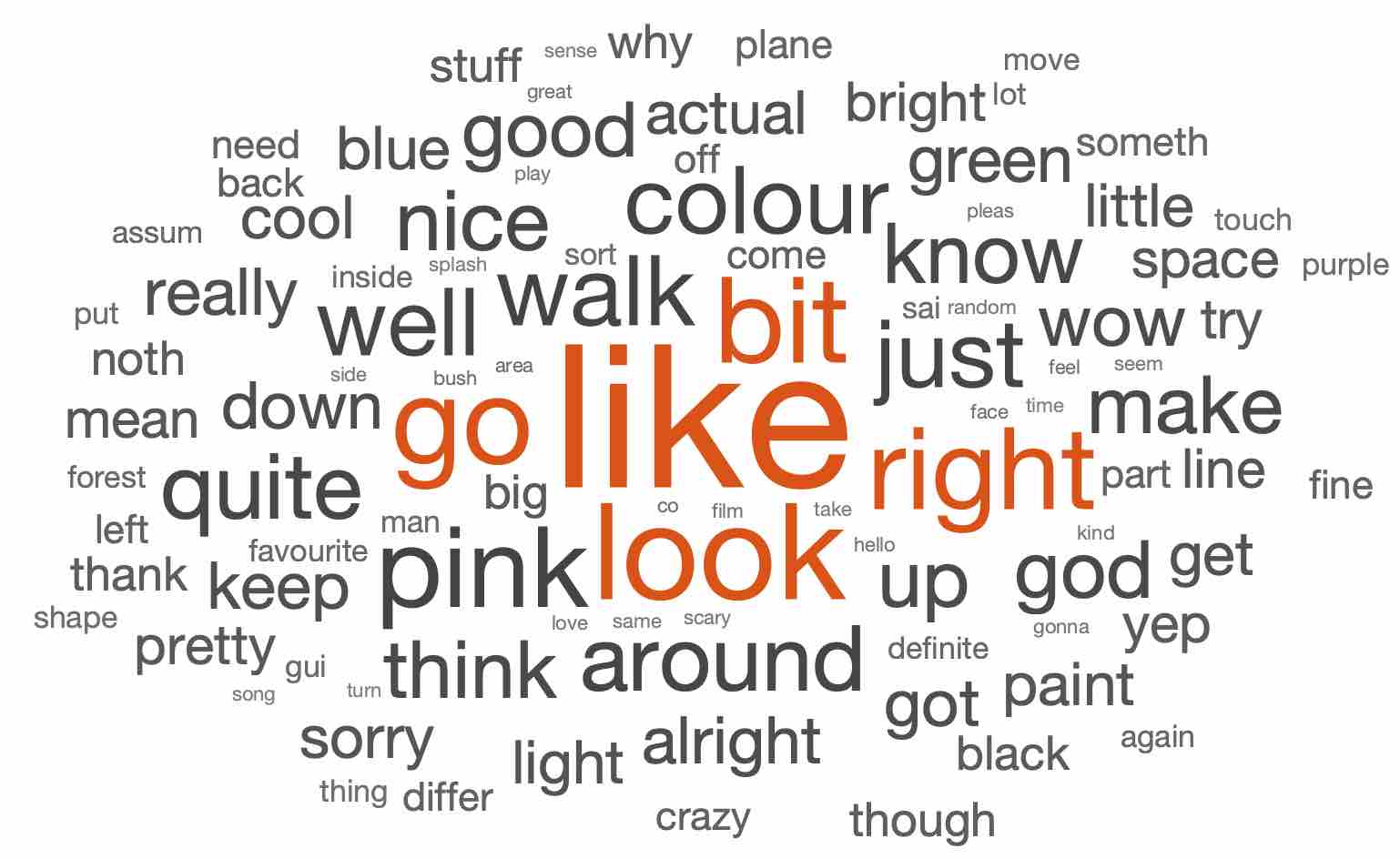}
\caption{Word cloud of audio recordings during experiment}
\label{fig:wordcloud}      
\end{figure}

Google Cloud Speech-to-Text API was used to automate the conversion from recorded audios to text. Figure \ref{fig:wordcloud} shows a word cloud of the verbal responses after the removal of stop words and filler words such as ``yeah". Most keywords are linked to general feelings towards the painting especially colours (``blue", ``green", ``purple"), lights (``bright"), scale (``big", ``little", ``space"), and direction (``right", ``down"). As young adults, participants also used words such as ``god", ``crazy", ``wow", ``scary" and ``cool", and ``like" to express their feelings towards the artwork. It was noticed during the experiment that a lot of the verbal responses were between the participants and the bystanders. In some cases, the ``cross-reality" conversations seemed to have encouraged some participants to explore the artwork as they became a performer. 

\section{Discussions}
\label{sec:discussions}

The user experiment resonate with what many artists already do while creating physical artwork. It’s common practice for artists to observe the audience/viewer/participant when encountering their artworks, to hide themselves from view and quietly watch the dynamics at play. In general, artworks go through many phases of testing starting at the artists’ studio, through various stages of experimentation, continued into more public spaces such project spaces or at screenings etc., which takes place long before any much grander launch is planned. This experiment keenly demonstrated how these kinds of quantitative processes of analysing responses could be a valuable tool to add to the process of VR content creation.

The post experiment interviews confirmed the artist author's intuition built upon their own encounters learning about how to make and navigate through an abstract VR painting. Those impressions being that here is a medium that has the potential to be immersive in a way currently beyond that of a traditional physical painting. These types of works could open up the field of painting to new audiences, such as the gaming community, with the possibility of expanding existing audiences by introducing them into new realms, that is Fine Art in VR.


Games engines and games design practices are likely to see an increasing adoption by the VR artists. Besides offering an environment to accommodate artwork, the games engines can empower content creators to exploit dynamic elements that respond to how the artwork are perceived by audiences. Game design theories such as the ``three Cs" (character, camera and control) and level design may also assist artists to choreograph audience interactions in alternate reality.

During the experiment, many participants reached their hands out and tried to stroke the artwork. In post experiment interviews, most participants welcomed the idea of ``touching" the artwork. While analysing the hand gesture data captured in the experiment was being carried out, the observations from the experiments indicate that multi-sensory experience may have a pivotal role in VR artwork. For abstract painting, this can lead to new experimental designs for how brushstrokes respond to user interactions with sound, visual deformation, and haptic feedback \cite{lecuyer2017playing,sonar2020closed}.

\section{Conclusions}
\label{sec:conclusions}

Alternate realities are becoming the new pathways for content creation, distribution and audience engagement in the creative communities. Understanding how audience explore and interact in new forms of media is essential to realise the full potential of alternate realities. Using a purposely built abstract VR painting and an experimentation system, our user experiment captured eye gaze and body movement patterns from 35 participants while interactions with the artwork occurred. The results can help expanding the knowledge base of user attention and interactions in VR especially in the context of fine art. Deep learning-based modelling showed how predictions of viewers' background can be made using behavioural data to potentially personalise the user experience. 



\bibliographystyle{spmpsci}      
\bibliography{references}   

\end{document}